\documentclass[MSNbibl,nameyear,dvips]{arxstspdf}
\usepackage{dcolumn}
\usepackage{graphicx}

\usepackage{flushend}
\usepackage{stfloats}

\def\bptnote#1{}

\volume{26}
\issue{3}
\pubyear{2011}
\firstpage{403}
\lastpage{422}
\doi{10.1214/11-STS360}

\makeatletter
\newcolumntype{d}[1]{D{.}{.}{#1}}
\newcommand{\expit}{\operatorname{expit}}
\newcommand{\logit}{\operatorname{logit}}
\newcommand{\ci}{\perp\!\!\!\perp}
\makeatother

\begin{document}
\begin{frontmatter}

\title{On Instrumental Variables Estimation of Causal Odds Ratios}
\runtitle{IV Estimation of Causal Odds Ratios}

\begin{aug}
\author{\fnms{Stijn} \snm{Vansteelandt}\corref{}\ead[label=e1]{stijn.vansteelandt@UGent.be}},
\author{\fnms{Jack} \snm{Bowden}\ead[label=e2]{jack.bowden@mrc-bsu.cam.ac.uk}},
\author{\fnms{Manoochehr} \snm{Babanezhad}\ead[label=e3]{m.babanezhad@gu.ac.ir}}
\and
\author{\fnms{Els} \snm{Goetghebeur}\ead[label=e4]{els.goetghebeur@UGent.be}}
\runauthor{Vansteelandt, Bowden, Babanezhad and Goetghebeur}

\affiliation{Ghent University, MRC Biostatistics Unit, Golestan
University and Ghent University}

\address{Stijn Vansteelandt is Associate Professor, Department of
Applied Mathematics and Computer Science, Ghent University, B-9000
Gent, Belgium \printead{e1}.
Jack Bowden is Research Fellow, MRC Biostatistics Unit, Robinson Way,
Cambridge, United Kingdom \printead{e2}.
Manoochehr Babanezhad is Assistant Professor, Department of Statistics,
Golestan University, Golestan, Gorgan, Iran \printead{e3}.
Els Goetghebeur is Professor, Department of Applied Mathematics and
Computer Science, Ghent University, B-9000 Gent, Belgium \printead{e4}.}
\end{aug}

% ABSTRACT
%
\begin{abstract}
Inference for causal effects can benefit from the availability of an
instrumental variable (IV) which, by definition, is associated with the
given exposure, but not with the outcome of interest other than through
a causal exposure effect. Estimation methods for instrumental variables
are now well established for continuous outcomes, but much less so for
dichotomous outcomes. In this article we review IV estimation of
so-called conditional causal odds ratios which express the effect of an
arbitrary exposure on a dichotomous outcome conditional on the exposure
level, instrumental variable and measured covariates. In addition, we
propose IV estimators of so-called marginal causal odds ratios which
express the effect of an arbitrary exposure on a dichotomous outcome at
the population level, and are therefore of greater public health
relevance. We explore interconnections between the different estimators
and support the results with extensive simulation studies and three
applications.
\end{abstract}

% KEYWORDS
%

\vspace*{-2pt}
\end{frontmatter}
%

%s1 ###
\section{Introduction}\label{sec1}\vspace*{-3pt}

Most causal analyses of observational data rely heavily on the
untestable assumption of no unmeasured confounders.\vadjust{\goodbreak} According to this
assumption, one has available all prognostic factors of the exposure
that are \textit{also} associated with the outcome \textit{other than
via} a possible exposure effect on outcome. Concerns about the validity
of this assumption plague observational data analyses and increase the
uncertainty surrounding many study results (Greenland, \citeyear
{Gre05}). This is
especially true in settings where the data analysis is based on
registry data or focuses on research questions different from those
conceived at the time of data collection. Substantial progress can
sometimes be made in settings where measurements are available on a
so-called instrumental variable (IV). This is a prognostic factor of
the exposure, which is \textit{not} associated with the outcome,
\textit
{except via} a possible exposure effect on outcome (Angrist, \citeyear{Ang90};
McClellan and Newhouse, \citeyear{McCMcNNew94}; Angrist, Imbens and
Rubin, \citeyear{AngImbRub96};
Hern\'{a}n and Robins, \citeyear{HerRob06}). An instrumental variable
$Z$ for the effect of
exposure $X$ on outcome $Y$ thus satisfies the following properties:
(a) $Z$ is associated with $X$; (b) $Z$ affects the outcome $Y$ only
through $X$ (i.e., often referred to as the exclusion restriction); (c)
the association between $Z$ and $Y$ is unconfounded\vadjust{\goodbreak} (i.e., often
referred to as the randomization assumption) (Hern\'{a}n and Robins,
\citeyear{HerRob06}). For instance, in the data analysis section, we
will estimate the
effect of Cox-2 treatment (versus nonselective NSAIDs) on
gastrointestinal bleeding, thereby allowing for the possibility of
unmeasured variables $U$ confounding the \mbox{association} between $X$ and
$Y$, by choosing the physician's prescribing preference for Cox-2
(versus nonselective NSAIDs) as an instrumental variable (Brookhart and
Schneeweiss, \citeyear{BroSch07}). Because this is associated with
Cox-2 treatment
[i.e., (a)], it would qualify as an IV if it were reasonable that the
physician's prescribing preference can only affect a patient's
gastrointestinal bleeding through his/her prescription [i.e., (b)] and
is not otherwise associated with that patient's gastrointestinal
bleeding [i.e., (c)]. Assumption~(b) could fail, however, if
preferential prescription of Cox-2 were correlated with other treatment
preferences that have their own impact on gastrointestinal bleeding;
the latter assumption could fail if patients with high risk of bleeding
are more often seen with physicians who prefer Cox-2 (Hern\'{a}n and
Robins, \citeyear{HerRob06}). In this article, we will more generally
assume that the
instrumental variables assumptions (a), (b) and~(c) hold conditional on
a (possibly empty) set of measured covariates $C$.

IVs have a long tradition in econometrics and are becoming increasingly
popular in biostatistics and epidemiology. This is partly because the
plausibility of a measured variable as an IV can sometimes be partially
justified on the basis of the study design or biological theory. For
instance, in randomized encouragement designs whereby, say, pregnant
women who smoke are randomly assigned to intensified encouragement to
quit smoking or not, randomization could qualify as an IV for assessing
the effects of smoking on low birth weight (Permutt and Hebel, \citeyear
{PerHeb89}),
since it guarantees the validity of IV assumption (c). The growing
success of IV methods in biostatistics and epidemiology can, however,
be mainly attributed to applications in genetic epidemiology (Smith and
Ebrahim, \citeyear{SmiEbr04}). Here, the random assortment of genes
transferred from
parents to offspring resembles the use of randomization in experiments
and is therefore often referred to as ``Mendelian randomization''
(Katan,
\citeyear
{Kat86}). Building on this idea, genetic variants may sometimes qualify as
an IV for estimating the relationship between a~genetically affected
exposure and a disease outcome, although violations of the necessary
conditions may occur (see Didelez and Sheehan, \citeyear{DidShe07}, and
Lawlor et al., \citeyear{Lawetal08}, for rigorous discussions).

Estimation methods for IVs are now well established for continuous
outcomes. The case of dichotomous outcomes has received more limited
attention. It turns out to be much harder because of the need for
additional modeling and because of difficulties to specify congenial
model parameterizations (see Sections \ref{exact-iv} and \ref
{secmarginal}). This paper therefore
combines different, scattered developments in the biostatistical,
epidemiological and econometric literature and aims to improve the
clarity and comparability of these developments by casting them within
a common causal language based on counterfactuals.

Traditional econometric approaches have their\break roots in structural
equations theory and have thereby largely focused on the estimation of
conditional cau\-sal effects, where rather than employing counterfactuals
to define causal effects, conditioning is made on all common causes,
$U$, of exposure $X$ and outcome~$Y$ (see Blundell and Powell, \citeyear
{BluPow03},
for a review). By this conditioning, one can assign a causal
interpretation to association measures such as
\[
\frac{\operatorname{odds}(Y=1|X=x+1,C,U)}{\operatorname{odds}
(Y=1|X=x,C,U)}.
\]
This can be seen by noting that this odds ratio measure can---under a
consistency assumption that $Y=Y(x)$ if $X=x$---equivalently be written
as (Pearl, \citeyear{Pea95})
%
%e1 ###
%
\begin{equation}\label{coru}
\frac{\operatorname{odds}\{Y(x+1)=1|C,U\}}{\operatorname{odds}\{
Y(x)=1|C,U\}},
\end{equation}
where $Y(x)$ denotes the (possibly) counterfactual outcome following an
intervention setting $X$ at the exposure level $x$ and where for any
$V,W$, $\operatorname{odds}(W=1|V)\equiv\mathrm{P}(W=1|V)/\mathrm{P}
(W=0|V)$. Effect measure~(\ref{coru}) thus compares the odds of
``success'' if the exposu\-re~$X$ were uniformly set to $x+1$ versus $x$
within~stra\-ta of $C$ and $U$. Because $U$ is unmeasured, these~stra\-ta
are not identified, which makes (\ref{coru}) less appealing as an
effect measure and of limited use for policy making. Its interpretation
is especially hindered in view of noncollapsibility of the odds ratio
(Greenland, Robins and Pearl, \citeyear{GreRobPea99}), following which
the magnitude of
conditional odds ratios chan\-ges with the conditioning sets, even in the
absence of confounding or effect modification. Similar limitations are
inherent to the so-called treatment effect on the treated at the IV
level $z$ of exposure $x$ (Tan, \citeyear{Tan10}),
%
%e2 ###
%
\begin{equation}\label{tan}
\frac{\operatorname{odds}\{Y(x)=1|X(z)=x\}}{\operatorname{odds}\{
Y(0)=1|X(z)=x\}},
\end{equation}
and to so-called local or principal stratification causal odds ratios
(Hirano et al., \citeyear{Hiretal00}; Frangakis\vadjust{\goodbreak} and Rubin \citeyear
{FraRub02}; Abadie, \citeyear{Aba03}; Clarke
and Windmeyer, \citeyear{ClaWin09}; see Bowden et al., \citeyear
{Bowetal}, for a review). For a
dichotomous instrumental variable $Z$ and dichotomous exposure $X$
taking values 0 and 1, the latter measure the association between
instrumental variable and outcome within the nonidentifiable principal
stratum of subjects for whom an increase in the instrumental variable
induces an increase in the exposure; that is,
%
%e3 ###
%
\begin{equation}\label{prinstrat}
\frac{\operatorname{odds}\{Y(1)=1|X(1)> X(0),C\}}{\operatorname
{odds}\{Y(0)=1|X(1)> X(0),C\}}.
\end{equation}
Inference for principal stratification causal odds ratios is also more
rigid in the sense of having no flexible extensions to more general
settings involving continuous instruments and exposures. While
dichotomization of the instrument and/or exposure is often employed in
view of this, it not only implies a loss of information, but may also
induce a violation of the exclusion restriction and may make the
relevance of the principal stratum ``$X(1)> X(0)$'' become dubious (see
Pearl, \citeyear{Pea11}, for further discussion of these issues).

In view of the aforementioned limitations, our attention in this
article will focus on causal effects which are defined within
identifiable subsets of the population. Special attention will be given
to the conditional causal odds ratio (Robins, \citeyear{Rob00};
Vansteelandt and Goetghebeur, \citeyear{VanGoe03};
Robins and Rotnitzky, \citeyear{RobRot04}), which we define as
%
%e4 ###
%
\begin{equation}\label{ccor}
\frac{\operatorname{odds}(Y=1|X,Z,C)}{\operatorname{odds}\{
Y(0)=1|X,Z,C\}}.
\end{equation}
It expresses the effect of setting the exposure to zero within
subgroups defined by the observed exposure level $X$, instrumental
variables $Z$ and covariates $C$. In the special case where $X$ is a
dichotomous treatment variable, taking the value 1 for treatment and~0
for no treatment, (\ref{ccor}) evaluated at $X=1$, that is,
\[
\frac{\operatorname{odds}\{Y(1)=1|X=1,Z,C\}}{\operatorname{odds}\{
Y(0)=1|X=1,Z,C\}}
\]
is sometimes referred to as the treatment effect in the treated who are
observed to have IV level $Z$ (Hern\'{a}n and Robins, \citeyear
{HerRob06}; Robins,
VanderWeele and Richardson, \citeyear{RobinsVanderWeele06}; Didelez, Meng and Sheehan,
\citeyear{DidMenShe10};
Tan, \citeyear{Tan10}). Conditional causal odds ratios would be of
special interest if
the goal of the study were to examine the impact of setting the
exposure to zero for those with a given exposure level $X$, for
example, to examine the impact of preventing nosocomial\vadjust{\goodbreak} infection
within those who acquired it (Vansteelandt et~al., \citeyear{Vanetal09}).

While the comparison in (\ref{ccor}) could alternatively be expressed
as a risk difference or relative risk, our focus throughout will be
limited to odds ratios because models for other association measures do
not guarantee probabilities within the unit interval, and might not be
applicable under case--control sampling (Bowden and Vansteelandt,
\citeyear{BowVan11}).
We refer the interested reader to Robins (\citeyear{Rob94}) and Mullahy
(\citeyear
{Mul97}) for
inference on the conditional relative risk
%
%e5 ###
%
\begin{equation}\label{rr}
\frac{\mathrm{P}(Y=1|X,Z,C)}{\mathrm{P}\{Y(0)=1|X,Z,C\}},
\end{equation}
and to van der Laan, Hubbard and Jewell (\citeyear{vanHubJew07}) for
inference on the
so-called switch relative risk, which is defined as (\ref{rr}) for
subjects with values $(X,Z,C)$ for which $\mathrm{P}(Y=1|X,Z,C)\leq
\mathrm{P}\{Y(0)=1|X,Z,C\}$ and as
\[
\frac{\mathrm{P}(Y=0|X,Z,C)}{\mathrm{P}\{Y(0)=0|X,Z,C\}},
\]
for all remaining subjects. The latter causal effect parameter is more
difficult to interpret, but has the advantage that models for the
switch relative risk, unlike models for (\ref{rr}), guarantee
probabilities within the unit interval.

For policy making, the interest lies more usually in
population-averaged or marginal effect measures (Greenland, \citeyear{Gre87};
Stock, \citeyear{Sto89}) such as
%
%e6 ###
%
\begin{eqnarray}\label{or1}
&&\frac{\operatorname{odds}\{Y(x+1)=1\}}{\operatorname{odds}\{
Y(x)=1\}},
\end{eqnarray}
where $x$ is a user-specified reference level, or
%
%e7 ###
%
\begin{eqnarray}\label{or2}
&&\frac{\operatorname{odds}\{Y(X+1)=1\}}{\operatorname{odds}\{
Y(X)=1\}} \quad\mbox{or}
\nonumber
\\[-8pt]
\\[-8pt]
\nonumber
&&\frac{\operatorname{odds}\{Y(1.1 \times
X)=1\}}{\operatorname{odds}\{Y(X)=1\}}.
\end{eqnarray}
Here, (\ref{or1}) evaluates the effect of changing the exposure from
level $x$ to $x+1$ uniformly in the population. It thus reflects the
effect that would have been estimated had an ideal randomized
controlled trial (i.e., with 100\% compliance) in fact been possible,
randomizing subjects over exposure level $x$ versus $x+1$. In contrast,
the effect measures in (\ref{or2}) allow for natural variation in the
exposure between subjects by expressing the effect of an absolute or
relative increase in the observed exposure. This may ultimately be of
most interest in many observational studies, considering that many
public\vadjust{\goodbreak} health interventions would target a change in exposure level
(e.g., diet, BMI, physical exercise, \ldots), starting from some natural,
subject-specific exposure level $X$.

We review estimation of the conditional causal odds ratio (\ref{ccor})
in Section \ref{sec2}. By casting different developments within the
same causal
framework based on counterfactuals, new insights into their
interconnections will be developed. We propose novel estimators of the
marginal causal odds ratios given in (\ref{or1}) and (\ref{or2}) in
Section \ref{secmarginal}, as well as for the corresponding effect
measures expressed
as risk differences or relative risks. Extensive simulation studies are
reported in Section \ref{Sim} and an evaluation on 3 data sets is given in
Section \ref{dataanalysis}.

%s2 ###
\section{IV Estimation of the Conditional Causal Odds Ratio}\label{sec2}

Identification of the conditional causal odds ratio (\ref{ccor}) is
studied in detail in Robins and Rotnitzky (\citeyear{RobRot04}) and
Vansteelandt and
Goetghebeur (\citeyear{VanGoe05}),
who find that---as for other IV estimators (Hernan
and Robins, \citeyear{HerRob06})---parametric restrictions are required
in addition to
the standard instrumental variables assumptions. In particular,
nonlinear exposure effects and modification of the exposure effect by
the instrumental variable are not nonparametrically identified. We will
therefore consider estimation of the conditional causal odds ratio
under so-called logistic structural mean models (Robins, \citeyear{Rob00};
Van\-steelandt and Goetghebeur, \citeyear{VanGoe03}; Robins and
Rotnitzky, \citeyear{RobRot04}), which
impose parametric restrictions on the conditional causal odds ratio
(\ref{ccor}). In particular, these models postulate that the exposure
effect is linear in the exposure on the conditional log odds ratio
scale, and independent of the instrumental variable, in the sense that
%
%e8 ###
%
\begin{equation}\label{VGmodel}
\frac{\operatorname{odds}(Y=1|X,Z,C)}{\operatorname{odds}\{
Y(0)=1|X,Z,C\}
}=\exp\{m(C;\psi^*)X\},\hspace*{-30pt}
\end{equation}
where $m(C;\psi)$ is a known function (e.g., $\psi_0+\psi_1C$), smooth
(i.e., with continuous first-order derivatives) in~$\psi$, and $\psi^*$
is an unknown finite dimensional parameter. In the absence of
covariates, this gives rise to a relatively simple model of the form
%
%e9 ###
%
\begin{equation}\label{VGmodelsimple}
\frac{\operatorname{odds}(Y=1|X,Z)}{\operatorname{odds}\{Y(0)=1|X,Z\}
}=\exp(\psi^*X).
\end{equation}
The assumption that the exposure effect is not modified by the IV
substitutes the monotonicity assumption [that $X(z)\,{\geq}\,X(z')$ if
$z\,{\geq}\,z'$] (Hernan and Robins, \citeyear{HerRob06}) which is commonly
adopted in the
principal stratification approach. In spite of the randomization
assumption [cf. IV assumption (c)], it may be violated \mbox{because}
subjects with exposure level $X$ are not exchangeable over levels of
the IV, so that they might in particular experience different effects.
The additional assumption of a linear exposure effect is only relevant
for exposures that take on more than two levels. It must be cautiously
interpreted because the conditional causal odds ratio (\ref{ccor})
expresses effects for differently exposed subgroups which may not be
exchangeable. Both these assumptions are critical because they are
empirically unverifiable.\break Vansteelandt and
Goetghebeur (\citeyear{VanGoe05}) assess
the sensitivity of the conditional causal odds ratio estimator to
violation of the linearity assumption and note that, under violation of
the linearity assumption, the estimator can still yield a meaningful
first order approximation. In the remainder of this work, we will
assume that model (\ref{VGmodel}) is correctly specified.

%s2.1 ###
\subsection{Approximate Estimation}
\label{approx-iv}

Approximate IV estimators of the conditional cau\-sal odds ratio can be
obtained by averaging over the observed exposure values in model (\ref
{VGmodel}) using the following approximations:
%
%e11 ###
%e10 ###
%
\begin{eqnarray}
\label{approx1}&&\mathrm{E}\{\logit\mathrm{E}(Y|X,Z,C)|Z,C\}
\nonumber
\\[-8pt]
\\[-8pt]
\nonumber
&&\quad\approx\logit\mathrm{E}
(Y|Z,C),\\
\label{approx2}&&\mathrm{E}[\logit\mathrm{E}\{Y(0)|X,Z,C\}|Z,C]
\nonumber
\\[-8pt]
\\[-8pt]
\nonumber
&&\qquad\approx\logit\mathrm{E}\{Y(0)|Z,C\}.
\end{eqnarray}
This together with the logistic structural mean mo\-del~(\ref
{VGmodel}) implies
\begin{eqnarray}\label{ttm}
&&\logit\mathrm{E}(Y|Z,C)\nonumber\\
&&\quad\approx\logit\mathrm{E}\{Y(0)|Z,C\}
+m(C;\psi^*)
\mathrm{E}(X|Z,C)\\
&&\quad=\logit\mathrm{E}\{Y(0)|C\}+m(C;\psi^*) \mathrm
{E}(X|Z,C),\nonumber
\end{eqnarray}
upon noting that the combined IV assumptions (b) and~(c), conditional
on $C$, imply $Y(x)\ci Z|C$ for all~$x$.
It follows that approximate IV estimators of the conditional causal
odds ratio can be obtained via the following two-stage approach:
\begin{enumerate}
\item Estimate the expected exposure in function of the IV and
covariates by fitting an appropriate regression model. Let the
predicted exposure be $\hat{X}\equiv\hat{\mathrm{E}}(X|Z,C)$.
\item Regress the outcome on covariates $C$ and on $m(C;\allowbreak\psi)\hat{X}$
through\vadjust{\goodbreak} standard logistic regression to obtain an estimate of $\psi^*$.
In the absence of covariates, this involves fitting a logistic
regression model of the form
\begin{equation}\label{stdest}\logit\mathrm{E}(Y|Z)=\omega+\psi
\hat{X}.
\end{equation}
When, furthermore, the IV is dichotomous, it follows from (\ref{ttm})
that
%
%e12 ###
%
\begin{eqnarray}\label{TTM-estimator}
\operatorname{OR}_{Y|Z}&\equiv&\frac{\operatorname
{odds}(Y=1|Z=1)}{\operatorname{odds}(Y=1|Z=0)}
\nonumber
\\[-8pt]
\\[-8pt]
\nonumber
&\approx&\exp(\psi^*)^{\Delta_{X|Z}},
\end{eqnarray}
where $\Delta_{X|Z}\equiv\mathrm{E}(X|Z=1)-\mathrm{E}(X|Z=0)$, so\break
that $\psi^*$ can be
estimated as $\log{\hat{\operatorname{OR}}_{Y|Z}}/\hat{\Delta}_{X|Z}$.
\end{enumerate}
The estimator obtained using the above two-stage approach is referred
to as the standard IV estimator in Palmer et al. (\citeyear
{Paletal08}), a Wald-type
estimator in Didelez, Meng and Sheehan (\citeyear{DidMenShe10}) and the
2-stage logistic
approach in Rassen et al. (\citeyear{Rasetal09}). It is commonly employed
in the
analysis of Mendelian randomization studies (Thompson et al., \citeyear
{ThoTobMin};
Palmer et al., \citeyear{Paletal08}), where it is typically viewed as
an approximate
estimator of the conditional causal odds ratio (\ref{coru}). Our
alternative development shows that it can also be viewed as an
approximate estimator of the conditional causal odds ratio (\ref
{ccor}). To gain insight into the adequacy of the approximations (\ref
{approx1}) and (\ref{approx2}), suppose for simplicity that there are
no covariates, that the exposure has a normal distribution with
constant variance $\sigma^2_x$ conditional on $Z$, that $\logit\mathrm{E}
(Y|X,Z)=\beta_0+\beta_xX+\beta_zZ$ and that $m(C;\allowbreak\psi)=\psi$. Then it
is easily shown, using results in Zeger and Liang (\citeyear
{ZegLiaAlb88}), that
\begin{eqnarray*}
\logit\mathrm{E}(Y|Z)&\approx&\beta_0\{\beta^{2}_{x}\sigma_x^2\}+\beta_x\{\beta^{2}_{x}\sigma_x^2\}\mathrm{E}(X|Z)\\
&&{}+\beta_z\{\beta^{2}_{x}\sigma_x^2\}Z,\\
\logit\mathrm{E}\{Y(0)|Z\}&\approx&\beta_0\{(\beta_{x}-\psi^*)^{2}\sigma_x^2\}\\
&&+(\beta_{x}\,{-}\,\psi^*)\{(\beta_{x}\,{-}\,\psi^*)^{2}\sigma_x^2\}\mathrm{E}(X|Z)\\
&&+\beta_z\{(\beta_{x}-\psi^*)^{2}\sigma_x^2\}Z,
\end{eqnarray*}
where for any parameter $\beta$ and variance component $\sigma^2$, we
define $\beta\{\sigma^{2}\} = \beta(c^{2}\sigma^{2} +
1)^{-{1}/{2}}$ with $c= 16\sqrt{3}/15\pi$. It can relatively easily
be deduced from these expressions and the fact that $\mathrm{E}\{
Y(0)|Z\}=\mathrm{E}\{
Y(0)\}$ that
\[
\logit\mathrm{E}(Y|Z)\approx\beta_0'+\frac{\psi^*}{\sqrt
{c^2\beta_x^2\sigma
_x^2+1}}\mathrm{E}(X|Z),
\]
for some $\beta_0'$, suggesting increasing bias with increasing
association between $X$ and $Y$ (given $Z$) and with increasing
residual variance in $X$ (given $Z$). This is true except\vadjust{\goodbreak} at the null
hypothesis of no causal effect because $Y\ci Z$ at the null hypothesis
so that the usual maximum likelihood estimator of $\psi$ in model~(\ref{stdest})
will then converge to 0 in probability. Further, note that
the standard IV estimator requires correct specification of the first
stage regression model for the expected exposure (Didelez, Meng and
Sheehan, \citeyear{DidMenShe10};
Rassen et al., \citeyear{Rasetal09}; Henneman, van der Laan and Hubbard,
\citeyear{HenvanHub}). In spite of its approximate nature, the standard
IV estimator
continues to be much used in Mendelian randomization studies because of
its simplicity, because it can be used in meta-analyses of summary
statistics, even when information on $\operatorname{OR}_{Y|Z}$ and
$\Delta_{X|Z}$ is
obtained from different studies (Minelli et al., \citeyear{Minetal04};
Smith et al.,
\citeyear{Smietal05}; Bowden et al., \citeyear{BowThoBur06}), and
because the underlying principle
extends to case--control studies when the first stage regression is
evaluated on the controls and the disease prevalence is low
(Smith et~al., \citeyear{Smietal05}; Bowden and Vansteelandt, \citeyear
{BowVan11}). For relative risk
estimators, the resulting bias due to basing the first stage regression
on controls rather than a random population sample amounts to the
difference between the log relative risk and the log odds ratio between
$Y$ and $Z$, inflated by the reciprocal of the exposure distortion
$\Delta_{X|Z}$ (Bowden and Vansteelandt, \citeyear{BowVan11}).

The bias of the standard IV estimator can someti\-mes be attenuated by
including the first-stage resi\-dual $R\equiv X-\hat{X}$ as an additional
regressor to $\hat{X}$~in model (\ref{stdest}). This is known as the
control functions~ap\-proach in econometrics (Smith and Blundell,
\citeyear{SmiBlu86}; Rivers and Vuong, \citeyear{RivVuo88}) and
has also been conside\-red in the
biostatistical literature on noncompliance adjustment (Nagelkerke et
al., \citeyear{Nagetal00}) and Mendelian randomization (Palmer et al.,
\citeyear{Paletal08}). A~control
function refers to a random variable conditioning on which renders the
exposure independent of the unmeasured variables that confound the
association between exposure and outcome. Intuitively, the regression
residual $R$ may apply as a control function because it captures (part
of) those confounders. In particular, let us summarize (without loss of
generality) all confounders of the exposure effect into a scalar
measurement $U$. Assume that the contributions of the instrument $Z$
and confounder $U$ are additive in the sense that $X=h(Z)+U$ for some
function $h$. Suppose for simplicity that there are no covariates and
that the conditional mean $\mathrm{E}(X|Z)$ is known so that $\hat{X}=h(Z)$
(here we use that $U\ci Z$, as implied by the IV assumptions). Then
\mbox{$R=U$} so that a (correctly specified) logistic regression
of~$Y$ on $X$\vadjust{\goodbreak}
and $R$ (or, equivalently, $\hat{X}$ and $R$) will yield a~consistent
estimator of the conditional causal odds ratio (\ref{coru}), which is
here identical to (\ref{ccor}) because $U$ is completely determined by
$X$ and $Z$. More generally, following the lines of Smith and Blundell
(\citeyear{SmiBlu86}),
\mbox{assume} that $X=h(Z)+V$, $U=\tilde{\beta}^*_1 V+\varepsilon
$, where~$\varepsilon$ follows a standard logistic distribution, and that\break $Y(x)=1$
if and only if $\tilde{\beta}^*_0+{\psi}^*x+U>0$ for some~$\tilde{\beta}^*_0$, $\tilde{\beta}^*_1$.
Then it also follows that $Y=1$ if and only
if $\varepsilon>-\tilde{\beta}^*_0-{\psi}^*X-\tilde{\beta}^*_1 V$,
from which
\[
\logit\mathrm{E}(Y|X,V)=\tilde{\beta}^*_0+{\psi}^*X+\tilde
{\beta}^*_1V.
\]
Upon substituting $V$ with the estimated regression residual $R$, one
obtains an estimator $\exp(\hat{\psi})$ which consistently estimates
the conditional causal odds ratio~(\ref{coru}). In the \hyperref
[app]{Appendix} we
demonstrate that this is also a consistent estimator of the conditional
causal odds ratio (\ref{ccor}) when the exposure is normally
distributed with constant variance, conditional on the instrument, but
not necessarily otherwise. Standard error calculation for the standard
and adjusted IV estimators is also detailed in the \hyperref[app]{Appendix}.

Over recent years, semiparametric analogs to the adjusted IV approach
have been developed in the econometrics literature to alleviate
concerns about model misspecification. Blundell and Powell (\citeyear
{BluPow04}) and
Rothe (\citeyear{Rot09}), for instance, avoid parametric restrictions
on the
conditional expectations $\mathrm{E}(X|Z,C)$ and $\mathrm{E}(Y|X,Z,C)$
(and, in
particular, on the distribution of $\varepsilon$) by using kernel
regression estimators and semiparametric maximum likelihood estimation,
respectively.
Imbens and Newey (\citeyear{ImbNew09}) allow for the contributions of
the instrument
$Z$ and confounder~$U$ on the exposure to be nonadditive by extending
the previous works to nonseparable exposure models of the form
$X=h(Z,U,C)$ for some function $h$. They show that the association
between exposure and outcome is unconfounded upon adjusting for
$R=F_{X|Z,C}(X|Z,C)$ as a control function, where $F_{X|Z,C}$ is the
conditional cumulative distribution function of $X$, given $Z$ and $C$.
To avoid parametric restrictions on the conditional expectations
$F_{X|Z,C}(X|Z,C)$ and $\mathrm{E}(Y|X,Z,C)$, they base inference on local
linear regression estimators.

$\!\!\!$A limitation of all these semiparametric approaches is that, by
avoiding assumptions on the distribution of~$\varepsilon$, the causal
parameter $\psi^*$ becomes difficult to interpret so that it may be
exclusively of interest for the calculation of marginal causal odds
ratios (see Section \ref{secmarginal}). A further limitation is that
all foregoing approaches require the exposure to be continuously
distributed (Rothe, \citeyear{Rot09}); some additionally require the IV
to be\vadjust{\goodbreak}
continuously distributed (Imbens and Newey, \citeyear{ImbNew09}). In
the next section
we review direct approaches to the estimation of the conditional causal
odds ratio~(\ref{ccor}) which do not rely on assumptions about the
exposure distribution.

%
%f1 ###
%
\begin{figure*}

\includegraphics{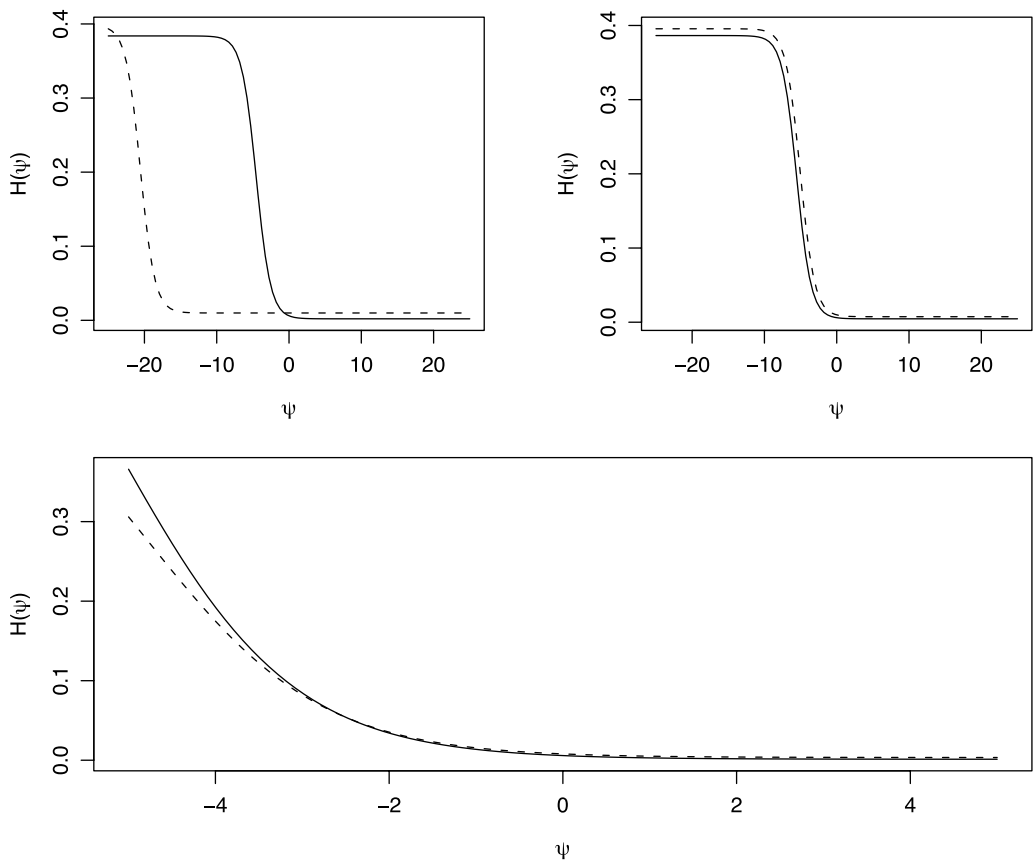}

\caption{Plot of the left- (solid) and right-hand side (dotted) of
expression (\protect\ref{closedform}) as a function of $\psi$. Top: simulated
data set [Right: with $\beta_4^*=0$ in model~(\protect\ref{assoc-example})];
Bottom: data set analyzed in Section~\protect\ref{dataanalysisbrookhart}.}
\label{fig1}
\vspace*{-3pt}
\end{figure*}

%s2.2 ###
\subsection{Consistent Estimation}\label{exact-iv}

$\!\!\!$Remember that, although $Y$ may well depend on~$Z$ (in the presence of
an exposure effect), the IV assumptions imply that $Y(0)\ci Z|C$.
Vansteelandt and Goetghebeur (\citeyear{VanGoe03}) make use of this to
obtain a~consistent estimator of $\psi^*$ in model (\ref{VGmodel}), which~is
chosen to make this independence happen. Because this is not possible
without making additional parametric modeling assumptions (Robins and
Rotnitzky, \citeyear{RobRot04}), they model the expected observed
outcome, conditional
on the exposure and IV, for example,
\begin{eqnarray}\label{assoc-example}
&&\logit\mathrm{P}(Y=1|X,Z,C)
\nonumber
\\[-8pt]
\\[-8pt]
\nonumber
&&\quad=\beta^*_0+\beta^*_1 X+\beta^*_2 Z+\beta^*_3 X
Z+\beta^*_4 C,
\end{eqnarray}
where $\beta^*_0,\beta^*_1,\beta^*_2,\beta^*_3$ and $\beta_4^*$ are
unknown scalar parameters.
More generally, one may postulate that
\begin{equation}
\logit\mathrm{E}(Y|X,Z,C)=m(X,Z,C;\beta^*), \label{association}
\end{equation}
where $m(X,Z,C;\beta)$ is a known function, smooth in~$\beta$,
and~$\beta^*$ is an unknown finite-dimensional parameter. An estimator
$\hat
{\beta}$ of $\beta^*$ can be obtained using standard methods (e.g.,
using maximum likelihood estimation).
Combining the causal model (\ref{VGmodel}) with the so-called
association model (\ref{association}) yields a prediction for the
counterfactual outcome $Y(0)$ for each subject which, for given $\psi
$, equals
\[
H(\psi,\hat{\beta})=\expit\{m(X,Z,C; \hat{\beta})-m(C;\psi) X\},
\]
where $\expit(a)\equiv{\exp(a)}/\{1+\exp(a)\}$.
Because\break $\mathrm{E}\{Y(0)|Z,C)=\mathrm{E}\{Y(0)|C\}$
under the IV assumptions, the value of $\psi^*$ can now be chosen as
the value $\psi$ which makes this mean independence happen, once $Y(0)$
is replaced by $H(\psi,\hat{\beta})$.
When there are no covariates and the instrument $Z$ is dichotomous,
taking the values 0 and 1, one thus chooses $\psi$ such that
\begin{equation}
\qquad\frac{\sum_i H_i(\psi,\hat{\beta}) Z_i}{\sum_i Z_i}=\frac{\sum_i
H_i(\psi,\hat{\beta})(1-Z_i)}{\sum_i(1-Z_i)}.
\label{closedform}
\end{equation}
When also the exposure is dichotomous, then mo\-del~(\ref{assoc-example})
is guaranteed to hold and a closed-form estimator is obtained, as given
in the \hyperref[app]{Appendix}.
In most cases, the solution to (\ref{closedform}) gives a unique
estimator of the causal odds ratio, although multiple or no solutions
are sometimes obtained when precision is limited due to small sample
size or the outcome mean being close to 0 or 1. This is illustrated in
Figure \ref{fig1},\vadjust{\goodbreak} which displays the left- and right-hand side of (\ref
{closedform}) in function of $\psi$ for 3 settings. The top 2 panels
are based on the same simulated data set. They show that 2 or no
solutions can be obtained for the same data set, depending on whether
the association model (\ref{association}) includes an interaction
between exposure and instrument (left panel) or not (right panel). The
bottom panel corresponds to the data analysis of Section \ref
{dataanalysisbrookhart}, where a single solution was obtained. Our
experience indicates that, when 2 solutions are obtained, one of them
corresponds to an effect size which is so large that it would be deemed
unrealistic [and correspondingly yield unrealistically small or large
values of $\mathrm{E}\{Y(0)\}$]. When no solutions are obtained, this can
sometimes be resolved by choosing a less parsimonious association model
(as in Figure~\ref{fig1}, top), but must be seen as an indication that
information is very limited. In the simulation experiments of Section
\ref{Sim}, a single solution was always obtained, but convergence of the
root-finding algorithm (nlm in R) was sometimes very dependent on the
choice of an adequate starting value.

For general instruments, a consistent point estimator of $\psi^*$ can
be found by solving unbiased estimating equation
\begin{eqnarray}\label{VGestimatingequation}
0&=&\sum_{i=1}^n [d(Z_i,C_i)-\mathrm{E}\{d(Z_i,C_i)|C_i\}
]
\nonumber
\\[-8pt]
\\[-8pt]
\nonumber
&&\hspace*{14pt}\cdot[H_i(\psi,\hat{\beta})-\mathrm{E}\{H_i(\psi,\hat
{\beta
})|C_i\}]
\end{eqnarray}
for $\psi$, where $d(Z_i,C_i)$ is an arbitrary function of~$Z_i$ and
$C_i$, for example, $d(Z_i,C_i)=Z_i$ (see Bowden and Vansteelandt,
\citeyear{BowVan11},
for choices that
yield a~semiparametric efficient estimator of $\psi^*$).
This thus leads to the following 2-stage approach:
\begin{enumerate}
\item First fit the association model (\ref{association}), for
instance, using maximum likelihood estimation, and obtain an estimator
$\hat{\beta}$ of\vadjust{\goodbreak} $\beta^*$;
\item Next, solve equation (\ref{VGestimatingequation}) to obtain an
estimator~$\hat{\psi}$ of $\psi^*$.
\end{enumerate}
Corresponding R-code is available from the first author's website
(\href{http://users.ugent.be/\textasciitilde
svsteela/Site/Welcome.html}{users.ugent.be/\textasciitilde svsteela}).
This app\-roach is extended in Tan (\citeyear{Tan10})\vadjust{\goodbreak} to enable estimation
of the
treatment effect on the treated at the IV level~$z$ of exposure $x$, as
defined in (\ref{tan}), thus avoiding conditioning on $C$.

In the \hyperref[app]{Appendix} we show that when the association model
includes an
additive term in
$d(Z_i,C_i)-\mathrm{E}\{d(Z_i,C_i)|C_i\}$ and is fitted using maximum
likeli\-hood estimation as in standard generalized linear~mo\-del software,
then its solution is robust to misspecifi\-cation of the association
model (\ref{association}) when \mbox{$\psi^*=0$}. This means that a consistent
estimator of $\psi^*=0$ is obtained, even when all models are
misspecified. In the absence of covariates and with $d(Z_i,C_i)=Z_i$
and $\mathrm{E}\{d(Z_i,C_i)|C_i\}=\sum_{j=1}^n Z_j/n$, this is
satisfied as soon as the association model includes an intercept and
main effect in $Z_i$ [as in model (\ref{assoc-example})]. The proposed
approach then yields a valid (Wald and score) test of the causal null
hypothesis that $\psi^*=0$, even when both models (\ref{VGmodel}) and
(\ref{association}) are misspecified. This property, which we refer to
as a ``local'' robustness property (Vansteelandt and Goetghebeur,
\citeyear{VanGoe03}),
also guarantees that estimators of the causal odds ratio will have
small bias under model misspecification when the true exposure effect
is close to, but not equal to, zero.

A drawback of the parameterization by Vansteelandt and Goetghebeur
(\citeyear{VanGoe03}) is that the
\mbox{association} model may be uncongenial with the causal
model. Specifically, given the observed data law $f(X,Z|C)$ and the
limiting value $\beta^*$ of $\hat{\beta}$, there may be no value of the
causal parameter $\psi$ for which $\mathrm{E}\{H(\psi,\beta^*)|Z,\allowbreak C\}
=\mathrm{E}\{H(\psi,\beta^*)|C\}$ over the entire support of $Z$ and $C$. In the \hyperref
[app]{Appendix},
we show that this may happen when parametric restrictions are imposed
on the main effect of the instrumental variable in the association
model (\ref{association}), along with its interaction with covariates
$C$, but not when that main effect is left unrestricted. It follows
that no congeniality problems arise in the common situation of
a~dichotomous instrument and no covariates, so long as a main effect of
the IV is included in the association model. This continues to be true
for categorical IVs with more than 2 levels when dummy regressors are
used for the instrument in the association model and there are no
covariates. For general IVs, one may consider generalized additive
association models which leave the main effect of the IV unrestricted
(apart from smoothness\vadjust{\goodbreak} restrictions).

Robins and Rotnitzky (\citeyear{RobRot04}) developed an alternative
approach for
estimation of $\psi^*$ in model (\ref{VGmodel}), which guarantees a
congenial parameterization by avoiding direct specification of an
association model. They parameterize instead the selection-bias\vadjust{\goodbreak} function
\begin{eqnarray}\label{selectionbias}
&&\logit\mathrm{E}\{Y(0)|X,Z,C\}\nonumber\\
&&\qquad{}-\logit\mathrm{E}\{Y(0)|X=0,Z,C\}
\\
&&\quad=q(X,Z,C; \eta^*),\nonumber
\end{eqnarray}
where $q(X,Z,C; \eta)$ is a known function satisfying $q(0,Z,C;\eta
)=0$, smooth in $\eta$, and $\eta^*$ is an unknown finite-dimensional
parameter. That $q(X,Z,C; \eta^*)$ encodes the degree of selection bias
can be seen because $q(X,Z,C; \eta^*)\,{=}\,0$ for all $X$ implies that
$\mathrm{E}\{Y(0)|\allowbreak X,Z, C\}=\mathrm{E}\{Y(0)| Z,C\}$ and thus implies
that the association between exposure and outcome [more precisely,
$Y(0)$] is unconfounded (conditional on $Z$ and $C$). Relying on a
parametric model for the conditional exposure distribution,
$f(X|Z,C)=f(X|Z,\allowbreak C;\alpha^*)$ (fitted using maximum likelihood
inference, for instance), their approach involves the following
iterative procedure. First, for each fixed $\psi$ (starting from an
initial value $\psi_0$), maximum likelihood estimators $\hat{\eta
}(\psi
)$ and $\hat{\omega}(\psi)$ are computed for the parameters $\eta^*$
and $\omega^*$ indexing the implied association model
%
%e13 ###
%
\begin{eqnarray}\label{impliedassoc}
&&\mathrm{P}(Y=1|X,Z,C;\psi,\eta^*,\omega^*)
\nonumber
\\
&&\quad=\expit\{m(C;\psi) X+ q(X,Z,C;
\eta^*)\\
&&\hspace*{70pt}\qquad{}+v(Z,C;\eta^*,\omega^*)\},\nonumber
\end{eqnarray}
where $v(Z,C;\eta^*,\omega^*)\equiv\logit\mathrm{E}\{Y(0)|X=0,Z,C\}
$ is the solution to the integral equation
%
%e15 ###
%e14 ###
%
\begin{eqnarray}\label{restr}
\logit\mathrm{E}\{Y(0)|C\}&=&t(C;\omega^*)\nonumber\\
&=& \int\expit\{q(X=x,Z,C; \eta^*)\\
&&\hspace*{39pt}\qquad{}+v(Z,C;\eta^*,\omega^*)\}\nonumber\\
&&\hspace*{8pt}\qquad{}\cdot f(X=x|Z,C;\alpha^*)\,dx,\nonumber
\end{eqnarray}
where $t(C;\omega)$ is a known function of $C$, smooth in~$\omega$, and
where $\omega^*$ is an unknown finite-dimensional parameter. For the
given estimators $\hat{\eta}(\psi)$ and $\hat{\omega}(\psi)$, an
estimator of $\psi$ is then obtained by solving a linear combination of
the estimating equations (\ref{VGestimatingequation})~and estimating
equations for the parameters indexing the association model (\ref
{impliedassoc}). Both these steps are~then~ite\-rated until convergence
of the estimator. In the \hyperref[app]{Ap-}\break\hyperref[app]{pendix} we suggest a somewhat simpler
strategy which, nonetheless, also involves solving integral equations.
Alternatively, one could focus on the switch relative risk of van der
Laan, Hubbard and Jewell (\citeyear{vanHubJew07}), introduced in
Section \ref{sec1}, to avoid the
uncongeniality problems associated with the odds ratio.

An advantage of the approach of Robins and Rotnitzky (\citeyear{RobRot04}) is that it
guarantees that $\mathrm{E}\{Y(0)|Z,\allowbreak C\}=\mathrm{E}\{Y(0)|C\}$ for
all $Z$ and $C$, although only under correct specification of the law
$f(X|Z,C)$. Under the approach of Vansteelandt and Goetghebeur
(\citeyear{VanGoe03}),
this is only guaranteed under congenial parameterizations as suggested
previously, but regardless of whether a model for the law $f(X|Z,C)$ is
(correctly) specified. A~further advantage is that it might possibly
give somewhat more efficient estimators by fully exploiting the a
priori knowledge that $\mathrm{E}\{Y(0)|Z,C\}=\mathrm{E}\{Y(0)|C\}$
to estimate unknown parameters [i.e., $v(Z,C)$] and by additionally
relying on a model for the exposure distribution. A drawback is that
the approach is computationally demanding, especially for continuous
IVs and/or in the presence of covariates, as it involves solving
integral equations for each $(Z,C)$ and this within each iteration of
the algorithm. In addition, standard error calculations are more
complex. A further drawback is that consistent estimation (away from
the null) requires correct specification of the conditional exposure
distribution $f(X|Z,C)$.

The estimation procedure for logistic structural mean models simplifies
when the logit link is replaced with the probit link and the exposure
is assumed to be normally distributed conditional on the instrumental
variable and covariates (with mean $\alpha^*_0+\alpha^*_1 Z+\alpha
_2^*C$ and constant standard deviation $\sigma^{*}$, where $\alpha
^*_0,\alpha^*_1,\sigma^{*}$ are unknown).
For instance, combining the probit structural mean model
\begin{eqnarray} \label{probitsmm}
\qquad&&\Phi^{-1}\{\mathrm{E}(Y|X,Z,C)\}-\Phi^{-1}\{ \mathrm{E}(Y(0)|
X,Z,C)\}
\nonumber
\\[-8pt]
\\[-8pt]
\nonumber
\qquad&&\quad= \phi^* X,
\end{eqnarray}
where $\Phi^{-1}$ is the probit link and $\phi^*$ is unknown, with the
probit association model
\begin{equation}\label{assoc-probit}
\qquad\Phi^{-1}\{\mathrm{E}(Y|X,Z,C)\}=\theta^*_0+\theta^*_1 X+\theta^*_2
Z+\theta^*_3 C,
\end{equation}
where $\theta^*_0,\theta^*_1,\theta^*_2$ are unknown, and averaging
over the exposure,
conditional on $Z$ and $C\!$ (see the \hyperref[app]{Appendix}), gives
\begin{eqnarray}\label{cumulative}
\qquad&&\mathrm{E}\{Y(0)|Z,C\}
\nonumber\\
\qquad&&\quad=\Phi\bigl\{\bigl(\theta_0^*+\theta_2^*
Z
\nonumber
\\[-8pt]
\\[-8pt]
\nonumber
\qquad&&\hspace*{4pt}\quad\qquad{} +(\theta_1^*-\phi^*)(\alpha^*_0+\alpha^*_1
Z+\alpha_2^*C)+\theta
_3^*C\bigr)\\
\qquad&&\hspace*{96pt}{}\cdot\bigl(\sqrt{1+(\theta_1^*-\phi^*)^2
\sigma^{2*}}\bigr)^{-1}\bigr\}.\nonumber
\end{eqnarray}
Because this does not depend on $Z$ under the IV assumptions, it
follows that $\theta_2^*=(\phi^*-\theta^*_1)\alpha^*_1$. Averaging over
the exposure in the association mo\-del~(\ref{assoc-probit}) and using
the previous identity, we obtain
\[
\mathrm{E}(Y|Z,C)=\Phi\biggl(\frac{\theta_0^*+\theta_1^*\alpha^*_0+\phi
^* \alpha
_1^* Z+\theta_3^*C}{\sqrt{1+\theta_1^{*2} \sigma^{2*}}}\biggr).
\]
This suggests regressing the outcome on the instrumental variable and
covariate using the probit regression model
\begin{equation}\label{probit-yz}\Phi^{-1}\{ \mathrm{E}(Y|Z,C)\}
=\lambda^*_0+\lambda^*_1
Z+\lambda_2^*C
\end{equation}
to obtain an estimate $\hat{\lambda}_1$ for the unknown regression
slope $\lambda_1^*$, and then estimating $\phi^*$ as
\begin{equation}
\hat{\phi}=\frac{\hat{\lambda}_1 \sqrt{1+\hat{\theta}_1^{2}
\hat{\sigma
}^2}}{\hat{\alpha}_1}. \label{Two-stage-estimatorII}
\end{equation}
We will refer to this estimator as the ``Probit-Normal SMM estimator''
throughout. It is related to the instrumental variables probit (Lee,
\citeyear{Lee81}) and the generalized two-stage simultaneous probit
(Amemiya, \citeyear{Ame78}),
both of which instead infer effect estimates conditional on the
unmeasured confounder $U$. When the outcome mean lies between 10\% and
90\%, the above estimator yields an approximate estimate of the causal
odds ratio through the identity $\exp(\psi^*)\approx\exp(\phi
^*/0.6071)$ (McCullagh and Nelder, \citeyear{McCNel83}). For dichotomous exposures,
related estimators can be obtained via probit structural equation
models that replace the linear regression model for $X_i$ in assumption
1 above, with a probit regression model (see, e.g., Rassen et al.,
\citeyear{Rasetal09}).

%s3 ###
\section{IV Estimation of the Marginal Causal Odds Ratio}\label{secmarginal}

We will now turn attention to the identification~of marginal causal
effects. Under linear structural models, these coincide with
conditional causal effects under typical assumptions (Hernan and
Robins, \citeyear{HerRob06}).
Consider, for instance, the extended linear structural mean model which
imposes the restriction
\[
\mathrm{E}\{Y-Y(x)|X,C,Z\}=m(C,x;\psi^*)(X-x)
\]
for \textit{each} feasible exposure level $x$, where $m(C,x;\psi)$ is a
known function (e.g., $\psi_0+\psi_1C+\psi_2x$), smooth in $\psi$, and
$\psi^*$ an unknown finite dimensional parameter. Then it follows from
the restriction
\begin{eqnarray*}
&&\mathrm{E}\{Y-m(C,x;\psi^*)(X-x)|C,Z\}\\
&&\quad=\mathrm{E}\{Y-m(C,x;\psi
^*)(X-x)|C\}
\end{eqnarray*}
for each $x$, that
\begin{eqnarray*}
&&\mathrm{E}\{Y-m(C,x;\psi^*)X|C,Z\}\\
&&\quad=\mathrm{E}\{Y-m(C,x;\psi^*)X|C\}
\end{eqnarray*}
for each $x$, and thus that $m(C,x;\psi^*)$ does not depend on $x$.
This then implies that the marginal cau\-sal effect equals
\[
\mathrm{E}\{Y(x^*)-Y(x)|C\}=m(C,0;\psi^*)(x^*-x).
\]
Unfortunately, this result does not extend to logistic structural mean
models, so that the conditional causal odds ratio corresponding to a
single reference exposure level (e.g., 0) does not uniquely map into
the marginal causal odds ratio.

Let us therefore assume that in addition to the association model (\ref
{association}),
the extended logistic structural mean model holds, which we define by
the restriction
%
%e16 ###
%
\begin{eqnarray}\label{GVGmodel}
&&\frac{\operatorname{odds}(Y=1|X,Z,C)}{\operatorname{odds}\{
Y(x)=1|X,Z,C\}
}
\nonumber
\\[-8pt]
\\[-8pt]
\nonumber
&&\quad=\exp\{m(C;\psi_x^*)(X-x)\},
\end{eqnarray}
for \textit{each} feasible exposure level $x$, where $m(C;\psi_x)$ is a
known function (e.g., $\psi_{x0}+\psi_{x1}C$), smooth in~$\psi_x$, and
$\psi_x^*$ an unknown finite-dimensional parameter.
The marginal causal odds ratio (\ref{or1}) can now be identified upon
noting that
\begin{eqnarray*}
&&\mathrm{P}\{Y(x)=1\}\\
&&\quad=\mathrm{E}[\expit\{m(X,Z,C;\beta^*)-m(C;\psi
_x^*)(X-x)\}]
\end{eqnarray*}
and the marginal causal odds ratio [(\ref{or2}), left] upon noting that
\begin{eqnarray*}
&&\mathrm{P}\{Y(X+1)=1\}\\
&&\quad=\mathrm{E}[\expit\{m(X,Z,C;\beta^*)+m(C;\psi
_{X+1}^*)\}].
\end{eqnarray*}
A consistent estimator of (\ref{or1}) is thus obtained by first
obtaining consistent estimators of $\beta^*,\psi_x^*$ and $\psi
_{x+1}^*$, using the strategy of the previous section, and then
calculating $\hat{p}_{x+1}(1-\hat{p}_x)/\{\hat{p}_x(1-\hat
{p}_{x+1})\}
$, where for given $x$
\begin{eqnarray*}
\hat{p}_x&=&n^{-1}\sum_{i=1}^n \expit\{m(X_i,Z_i,C_i;\hat{\beta
})\\
&&\hspace*{36pt}\qquad{}-m(C_i;\hat{\psi}_x)(X_i-x)\}.
\end{eqnarray*}
A consistent estimator of [(\ref{or2}), left] is obtained by first
obtaining consistent estimators of $\beta^*$ and $\psi_{x+1}^*$ for
each observed value $X_i$ for $x$ using the strategy of the previous
section, and then calculating $\hat{p}_{X+1}(1-\hat{p}_X)/\{\hat
{p}_X(1-\hat{p}_{X+1})\}$, where
\begin{eqnarray*}
\hat{p}_X&=&n^{-1}\sum_{i=1}^n Y_i,\\
\hat{p}_{X+1}&=&n^{-1}\sum_{i=1}^n \expit\{m(X_i,Z_i,C_i;\hat{\beta
})\\
&&\hspace*{36pt}\qquad{}+m(C_i;\hat{\psi}_{X_i+1})\}.
\end{eqnarray*}
Standard error calculations are reported in the \hyperref[app]{Ap-}\break\hyperref[app]{pendix}. Using the
above expressions, also estimators of the marginal risk difference
$\mathrm{P}\{
Y(x+1)=1\}-\mathrm{P}\{Y(x)=1\}$ or relative risk $\mathrm{P}\{Y(x+1)=1\}/\allowbreak\mathrm{P}\{Y(x)=1\}$
can straightforwardly be obtained.

A drawback of this strategy, which we discuss in the Appendix, is that
even when model (\ref{GVGmodel}) is congenial with the association
model (\ref{association}) for $x=0$ (or some other reference level), it
need not be a well-specified model for all $x$. We conjecture that when
this would happen, this may be partially detectable in the sense of
yielding estimating equations with no solution, as the uncongeniality
is then due to the nonexistence of a value of $\psi_x^*$ for some $x$
so that $\mathrm{E}\{Y(x)|Z,C\}=\mathrm{E}\{Y(x)|C\}$ for all $(Z,C)$.
As with other
causal models that are not guaranteed to be congenial (e.g., Petersen
et al., \citeyear{Petetal07}; Tan, \citeyear{Tan10}) and as confirmed in
simulation studies in the
next section, we believe this is unlikely to induce an important bias.
The concern for bias is further alleviated by the aforementioned local
robustness property, which continues to hold for extended logistic
structural mean models.

The idea of using conditional causal effect estimates as plug-in
estimates in inference for marginal effects has been advocated in the
biostatistical and epidemiological literature (see, e.g., Greenland,
\citeyear{Gre87}; Ten Have et al., \citeyear{HavJofCar03}) and is commonly
employed in the
econometrics literature (see, e.g., Blundell and Powell, \citeyear
{BluPow04}; Imbens and Newey, \citeyear{ImbNew09}), where
related proposals have been made starting from
a~semiparametric control functions approach. Alternative approaches
involve assuming that all confounders of the exposure effect can be
captured into a scalar variate $U$, which has an additive effect on the
outcome (Amemiya, \citeyear{Ame74}; Foster, \citeyear{Fos97}; Johnston
et al., \citeyear{Johetal08};
Rassen et
al., \citeyear{Rasetal09}) in the sense that
\begin{equation}\label{johnston}
\qquad\mathrm{E}(Y|X,C,U)=\expit(\beta_0^*+\tilde{\psi}^* X+\beta
_1^*C)+ U,
\end{equation}
where
$\beta^*_0,\beta^*_1,\tilde{\psi}^*$ are unknown and where
$\mathrm{E}(U|C)\,{=}\,0$; note that \mbox{$\mathrm{E}(U|X,C)\,{\ne}\,0$} when there is
confounding.
Because, for each $x$, $Y(x)\ci X|U,C$, model (\ref{johnston}) implies
the marginal structural model
\begin{eqnarray*}
\mathrm{E}\{Y(x)|C\}&=&\mathrm{E}[\mathrm{E}\{Y(x)|X=x,C,U\}|C
]\\
&=&\expit(\beta_0^*+\tilde{\psi}^* x+\beta_1^*C)
\end{eqnarray*}
considered by Henneman, van der Laan and \mbox{Hubbard} (\citeyear
{HenvanHub}). This clarifies
that $\exp(\tilde{\psi}^*)$ in model (\ref{johnston}) can be
interpreted as the marginal (i.e., population averaged) causal odds ratio
\[
\exp(\tilde{\psi}^*)=\frac{\operatorname{odds}\{Y(1)=1|C\}
}{\operatorname
{odds}\{Y(0)=1|C\}}.
\]
Using that $Z\ci U|C$ under the IV assumptions, an estimator $\hat
{\psi
}$ for $\psi^*$ can be obtained by solving the following unbiased
estimating equations:
%
%e17 ###
%
\begin{equation}\label{johnstonee}
0=\sum_{i=1}^n \pmatrix{
1 \cr Z_i \cr C_i}
\{ Y_i-\expit({\beta}_0+{\psi}X_i+{\beta}_1C_i)\}.\hspace*{-30pt}
\end{equation}
The marginal causal odds ratio (\ref{or1}) can be identified upon
noting that
\[
\mathrm{P}\{Y(x)=1\}=\mathrm{E}[\expit(\beta_0^*+\tilde{\psi}^* x+\beta
_1^*C)];
\]
it equals $\exp(\tilde{\psi}^*)$ when $C$ is empty. The marginal causal
odds ratio [(\ref{or2}), left] can be identified
upon noting that
\begin{eqnarray*}
&&\mathrm{P}\{Y(X+1)=1\}\\
&&\quad=\mathrm{E}[\expit\{\beta_0^*+\tilde{\psi}^*
(X+1)+\beta_1^*C\}].
\end{eqnarray*}
In the absence of covariates, it follows from the unbiasedness of the estimating functions at $\tilde
{\psi}^*=0$ that the resulting estimator
is (locally) robust against model misspecification at the null
hypothesis of no causal effect. However, it is not guaranteed to exist
and may be inconsistent for $\psi^*\ne0$ because the dichotomous
nature of the outcome imposes strong restrictions on the distribution
of $U$, which may be impossible to reconcile with the basic assumption
that $Z\ci U|C$ (Henneman, van der Laan and Hubbard, \citeyear{HenvanHub}).

%s4 ###
\section{Simulation Study}\label{Sim}

We conducted 5 simulation experiments, each with a sample size of 1,000
and with 1,000 simulation runs. As in Palmer et al. (\citeyear
{Paletal08}), the
instrumental variable $Z$ was generated in such a manner as to
represent the number of copies (0, 1 or 2) of a single bi-allelic SNP
in the Hardy--Weinberg equilibrium. The underlying allele frequency in
the population was assumed to be $p=0.3$, and so $Z$ was generated from
a multinomial distribution with cell probabilities (0.09, 0.42, 0.49).
The exposure $X$ was generated to be $N(Z,2)$ in simulation experiments
a, b and e, $Z+t_2$ in simulation experiment c and $\Gamma(Z,1)$ in
simulation experiment d [with $\Gamma(\cdot,\cdot)$ referring to the Gamma
distribution].\vadjust{\goodbreak} Finally, the outcome was generated to satisfy
\[
\mathrm{P}(Y=1|X,Z)=\operatorname{expit}(\beta_0+\beta_xX+\beta_zZ),
\]
where $\beta_0$ was fixed at different values to result in outcome
means of 0.05, 0.1, 0.25 and 0.5 and $\beta_x$ was chosen to yield
$Y(0)\ci Z$ under the logistic structural mean model (\ref
{VGmodelsimple}) with $\psi$ equaling 0 or 1. Finally, $\beta_z$ was
set to 1 in simulation experiments a and e, to 2 in simulation
experiments b and c and to $-2$ in simulation experiment d to correspond
to different degrees of unmeasured confounding. Indeed, note that the
conditional association $\beta_z$ between $Y(0)$ and $Z$, conditional
on $X$, is largely explained by the extent of unmeasured confounding.

Table \ref{tab1} compares the Wald estimator, the Adjusted IV estimator
and the
logistic structural mean model estimator of the conditional causal log
odds ratio. We do not report results for the semiparametric control
function approaches since these require the IV to be continuously
distributed (Imbens and Newey, \citeyear{ImbNew09}). Table \ref{tab1}
demonstrates that the
Wald estimator can have substantial bias when there is unmeasured
confounding of the exposure--outcome association (cf. experiment b). As
predicted by the theory, the adjusted IV estimator gives unbiased
estimators when the exposure has a symmetric distribution with constant
variance (cf. experiments a--c), conditional on the IV, but not when
the exposure distribution is skewed (cf. experiment d) or when an
exposure--IV interaction is ignored (cf. experiment~e). Note, in
particular, that the adjusted IV estimator is not locally robust to
model misspecification at the causal null hypothesis $\psi^*=0$,
despite the existence of an asymptotically distribution-free test. The
logistic SMM estimator is unbiased in all cases. It has slightly
increased variance relative to the Adjusted IV estimator when the
exposure is normally distributed, but reduced variance when the
exposure is $t$-distributed because of outlying exposure residuals (i.e.,
control functions) affecting the Adjusted IV estimator.

%
%t1 ###
%
\begin{table*}
\tabcolsep=0pt
\caption{Bias ($\times100$), empirical standard deviation ($\times
100$) (ESE), average sandwich standard error ($\times100$) (SSE) and
coverage of 95\% confidence intervals (Cov.) for the standard IV
estimator, the adjusted IV estimator and the logistic structural mean
model estimator of the log conditional causal odds ratio}\label{tab1}
\begin{tabular*}{\textwidth}{@{\extracolsep{4in minus 4in}}ld{1.2}cd{3.2}d{2.2}d{2.2}d{2.2}d{3.2}d{2.1}d{2.1}d{2.1}d{2.2}d{2.1}d{2.1}d{2.1}@{}}
\hline
& & &
\multicolumn{4}{c}{\textbf{Standard IV}} &
\multicolumn{4}{c}{\textbf{Adjusted IV}} &
\multicolumn{4}{c@{}}{\textbf{Logistic SMM}}\\
\ccline{4-7,8-11,12-15}\\[-8pt]
\multicolumn{1}{@{}l}{\textbf{Exp.}} &
\multicolumn{1}{c}{$\bolds{\mathrm{E}(Y)}$} &
\multicolumn{1}{c}{$\bolds{\psi}$} &
\multicolumn{1}{c}{\textbf{Bias}} &
\multicolumn{1}{c}{\textbf{ESE}} &
\multicolumn{1}{c}{\textbf{SSE}} &
\multicolumn{1}{c}{\textbf{Cov.}} &
\multicolumn{1}{c}{\textbf{Bias}} &
\multicolumn{1}{c}{\textbf{ESE}} &
\multicolumn{1}{c}{\textbf{SSE}} &
\multicolumn{1}{c}{\textbf{Cov.}} &
\multicolumn{1}{c}{\textbf{Bias}} &
\multicolumn{1}{c}{\textbf{ESE}} &
\multicolumn{1}{c}{\textbf{SSE}} &
\multicolumn{1}{c@{}}{\textbf{Cov.}}\\
\hline
a&0.1&0& 1.15 & 16.2 & 15.9 & 95.5 & 1.11 & 19.2 & 18.9 & 95.1 & 1.62 &
20.1 & 19.6 & 95.6\\
&0.05&1&3.82 & 30.8 & 30.4 & 96.1 & 3.92 & 30.8 & 30.5 & 96.0 & 5.31 &
33.0 & 32.2 & 96.2\\
&0.1&1&1.71 & 22.0 & 21.9 & 95.3 & 1.80 & 22.0 & 21.9 & 95.5 & 2.71 &
23.6 & 23.0 & 95.6\\
&0.25&1&0.68 & 15.0 & 15.0 & 95.5 & 0.77 & 15.0 & 15.1 & 95.6 & 1.24 &
15.8 & 15.7 & 95.3\\
&0.5&1&1.18 & 12.3 & 12.7 & 95.1 & 1.28 & 12.3 & 12.7 & 95.3 & 1.46 &
12.6 & 13.0 & 95.6\\[3pt]
b&0.1&0&1.28 & 15.7 & 15.9 & 95.1 & 1.31 & 24.8 & 25.1 & 95.5 & 2.86 &
28.3 & 28.3 & 95.9\\
&0.05&1&-7.12 & 31.1 & 27.9 & 88.9 & 4.38 & 34.4 & 33.4 & 95.3 & 6.63 &
38.7 & 37.3 & 95.1\\
&0.1&1&-13.5 & 22.1 & 18.9 & 80.1 & 2.69 & 25.4 & 25.7 & 95.3 & 4.37 &
29.0 & 28.9 & 95.2\\
&0.25&1&-21.9 & 15.3 & 11.6 & 49.2 & 1.84 & 19.8 & 20.1 & 95.1 & 2.76 &
22.1 & 22.2 & 95.8\\
&0.5&1&-26.0 & 13.2 & 8.89 & 26.5 & 1.28 & 18.0 & 18.3 & 95.4 & 1.65 &
19.3 & 19.4 & 95.4\\[3pt]
c&0.1&0&1.77 & 17.0 & 17.1 & 95.0 & 7.06 & 73.5 & 61.3 & 94.4 & 5.31 &
39.8 & 39.5 & 95.2\\
&0.05&1&-34.8 & 36.1 & 30.4 & 55.4 & 10.8 & 79.8 & 69.4 & 94.4 & 12.2 &
58.0 & 56.2 & 93.1\\
&0.1&1&-29.1 & 34.9 & 26.6 & 50.5 & 9.82 & 83.0 & 63.5 & 94.8 & 8.15 &
41.1 & 39.7 & 95.9\\
&0.25&1&-25.6 & 30.3 & 21.2 & 41.9 & 7.45 & 68.3 & 54.2 & 93.1 & 3.50 &
26.9 & 25.2 & 95.1\\
&0.5&1&-24.7 & 26.8 & 18.9 & 39.2 & 7.23 & 66.9 & 53.1 & 93.8 & 1.8 &
19.7 & 19.0 & 95.3\\[3pt]
d&0.1&0&0.08 & 15.6 & 15.8 & 95.3 & -56.2 & 25.6 & 26.2 & 40.8 & -1.03 &
28.6 & 28.5 & 94.0\\
&0.05&1&-48.0 & 24.1 & 26.2 & 51.7 & -91.8 & 47.0 & 43.5 & 42.4 & -1.09
& 40.0 & 34.1 & 87.7\\
&0.1&1&-55.8 & 15.8 & 19.0 & 14.3 & -83.4 & 32.3 & 31.9 & 22.3 & 1.16
& 33.5 & 31.6 & 88.2\\
&0.25&1&-65.2 & 9.87 & 13.1 & 0.00 & -61.8 & 23.3 & 23.1 & 21.2 & 1.59
& 26.8 & 27.0 & 94.0\\
&0.5&1&-72.8 & 8.53 & 10.8 & 0.00 & -27.0 & 19.9 & 20.3 & 76.3 & -0.07
& 27.3 & 28.5 & 95.2\\[3pt]
e&0.1&0&2.55 & 15.5 & 15.4 & 94.8 & 2.83 & 18.6 & 19.2 & 95.9 & 3.25 &
26.8 & 26.4 & 97.2\\
&0.05&1&-37.7 & 25.8 & 25.2 & 62.2 & -37.4 & 26.0 & 25.8 & 64.0 & 13.4
& 56.3 & 52.4 & 91.0\\
&0.1&1&-36.6 & 18.4 & 18.3 & 45.0 & -36.4 & 18.6 & 18.9 & 48.1 & 8.38
& 39.8 & 38.0 & 93.9\\
&0.25&1&-31.0 & 12.7 & 13.0 & 34.3 & -30.9 & 12.7 & 13.2 & 35.6 & 4.83
& 24.4 & 24.3 & 95.7\\
&0.5&1&-19.1 & 10.7 & 11.9 & 61.8 & -18.7 & 10.8 & 11.4 & 60.4 & 4.18 &
17.1 & 17.4 & 96.0\\
\hline
\end{tabular*}
\end{table*}

%
%t2 ###
%
\begin{table*}
\tabcolsep=0pt
\caption{Bias ($\times100$), empirical standard deviation ($\times
100$) (ESE), average sandwich standard error ($\times100$) (SSE) and
coverage of 95\% confidence intervals (Cov.) for the approximate and
exact estimators of the logarithm of (\protect\ref{or1}) (MLOR1) and the
logarithm of (\protect\ref{or2}) (leftmost) (MLOR2)}\label{tab2}
\begin{tabular*}{\textwidth}{@{\extracolsep{\fill
}}ld{1.2}cd{2.2}d{2.2}d{2.2}cd{2.2}d{2.2}d{2.2}c@{}}
\hline
& & & \multicolumn{4}{c}{\textbf{Approx. MLOR 1}} &
\multicolumn{4}{c@{}}{\textbf{MLOR 1}}\\
\ccline{4-7,8-11}\\[-8pt]
\multicolumn{1}{@{}l}{\textbf{Exp.}} & \multicolumn{1}{c}{$\bolds
{\mathrm{E}(Y)}$} & \multicolumn{1}{c}{$\bolds{\psi}$}& \multicolumn
{1}{c}{\textbf{Bias}} &
\multicolumn{1}{c}{\textbf{ESE}} &
\multicolumn{1}{c}{\textbf{SSE}} & \multicolumn{1}{c}{\textbf
{Cov.}} &
\multicolumn{1}{c}{\textbf{Bias}} & \multicolumn{1}{c}{\textbf
{ESE}} &
\multicolumn{1}{c}{\textbf{SSE}} & \multicolumn{1}{c@{}}{\textbf{Cov.}}
\\
\hline
a&0.1&0& -0.10 & 15.9 & 15.6 & 93.7 & -0.30 & 15.7 & 15.6 & 94.9\\
&0.05&1&-0.31 & 9.82 & 9.79 & 93.6 & -0.65 & 9.54 & 10.1 & 96.0\\
&0.1&1&-1.01 & 14.2 & 14.3 & 92.6 & -1.52 & 14.0 & 15.0 & 95.2\\
&0.25&1&0.04 & 6.50 & 6.51 & 94.7 & -0.16 & 6.31 & 6.52 & 95.6\\
&0.5&1&0.32 & 5.44 & 5.56 & 95.9 & 0.24 & 5.46 & 5.50 & 94.1\\[3pt]
b&0.1&0& -0.49 & 15.5 & 15.9 & 94.1 & -1.30 & 16.1 & 16.1 & 95.5\\
&0.05&1&-0.04 & 11.0 & 10.8 & 94.0 & -0.58 & 10.7 & 12.0 & 96.4\\
&0.1&1&0.23 & 8.29 & 8.35 & 94.2 & -0.24 & 7.89 & 8.90 & 96.3\\
&0.25&1&0.18 & 6.42 & 6.49 & 95.1 & -0.06 & 6.12 & 6.52 & 96.1\\
&0.5&1& 0.07 & 5.80 & 5.87 & 95.8 & -0.04 & 5.70 & 5.80 & 95.5\\[6pt]
& & & \multicolumn{4}{c}{\textbf{Approx. MLOR 2}} &
\multicolumn{4}{c}{\textbf{MLOR 2}}\\
\ccline{4-7,8-11}\\[-8pt]
& & & \multicolumn{1}{c}{\textbf{Bias}} & \multicolumn{1}{c}{\textbf
{ESE}} & \multicolumn{1}{c}{\textbf{SSE}} & \multicolumn
{1}{c}{\textbf
{Cov.}} &
\multicolumn{1}{c}{\textbf{Bias}} &
\multicolumn{1}{c}{\textbf{ESE}} & \multicolumn{1}{c}{\textbf{SSE}} &
\multicolumn{1}{c@{}}{\textbf{Cov.}}\\
\hline
a&0.1&0& 1.2 & 16.4 & 16.1 & 95.5 & 1.14 & 16.3 & 16.0 & 94.9\\
&0.05&1& 1.57 & 20.6 & 20.2 & 95.6 & 1.04 & 19.4 & 23.9 & 94.8\\
&0.1&1& 4.00 & 30.1 & 29.5 & 95.6 & 3.06 & 28.4 & 28.1 & 95.0\\
&0.25&1& 0.24 & 12.7 & 12.7 & 95.1 & 0.00 & 12.0 & 14.0 & 95.7\\
&0.5&1& 0.29 & 9.93 & 10.3 & 95.9 & 0.25 & 9.8 & 10.1 & 95.9\\[3pt]
b&0.1&0& 1.46 & 15.9 & 16.1 & 95.8 & 1.28 & 15.7 & 15.9 & 95.2\\
&0.05&1& 3.74 & 28.4 & 28.7 & 96.4 & 2.97 & 26.3 & 26.7 & 95.3\\
&0.1&1& 2.41 & 20.1 & 21.0 & 96.6 & 1.72 & 18.4 & 19.2 & 95.5\\
&0.25&1& 1.24 & 14.2 & 15.1 & 96.5 & 0.80 & 12.8 & 13.9 & 96.0\\
&0.5&1& 0.58 & 12.4 & 13.3 & 97.1 & 0.52 & 12.0 & 13.1 & 96.7\\
\hline
\end{tabular*}
%
% \end{sidewaystable}
\vspace*{12pt}
\end{table*}

Table \ref{tab2} compares the proposed estimators of the marginal log
odds ratio
(\ref{or1}) (labeled ``MLOR 1'') and~(\ref{or2}) (labeled ``MLOR 2''), as
well as the same estimators where, for computational convenience, $\hat
{\psi}_x$ is substituted with $\hat{\psi}_0$ for all $x$ (labeled
``Approx. MLOR 1'' and ``Approx. MLOR 2''). We do not report results on the
estimators obtained by solving~(\ref{johnstonee}) since they were doing
very poorly, often resulting in nonconvergence in over 80\% of the
simulation runs. Table \ref{tab2} demonstrates that the approximate estimators
perform adequately and much like the proposed estimators, although the
nominal coverage level is slightly better attained for the proposed
estimators. Given the good agreement, the results in Table \ref{tab3}
are based
on the computationally more attractive approximate estimators.
Interestingly, it reveals that the estimators of the marginal causal
log odds ratio have a much reduced variance relative to the three
considered estimators of the conditional causal log odds ratio. In
particular, highly efficient estimates are obtained for the marginal
causal log odds ratio (\ref{or1}) which we regard to be of most
interest in many practical applications, since it essentially expresses
the result that would be obtained in a randomized experiment.

%
%t3 ###
%
\begin{table*}
\tabcolsep=0pt
\caption{Bias ($\times100$), empirical standard deviation ($\times
100$) (ESE), average sandwich standard error ($\times100$) (ESE) and
coverage of 95\% confidence intervals (Cov.) for the logistic
structural mean model estimator of the log conditional causal odds
ratio (\protect\ref{ccor}), the approximate estimator of the logarithm
of (\protect\ref
{or1}) (MLOR1) and the logarithm of (\protect\ref{or2}) (leftmost)
(MLOR2)}\label{tab3}
\begin{tabular*}{\textwidth}{@{\extracolsep{\fill
}}ld{1.2}cd{2.2}d{2.1}d{2.1}d{2.1}d{2.2}d{2.2}d{2.2}d{2.1}d{2.2}d{2.2}d{2.2}d{2.1}@{}}
\hline
& & & \multicolumn{4}{c}{\textbf{Logistic SMM}} & \multicolumn
{4}{c}{\textbf{MLOR 1}} & \multicolumn{4}{c@{}}{\textbf{MLOR2}}\\
\ccline{4-7,8-11,12-15}\\[-8pt]
\multicolumn{1}{@{}l}{\textbf{Exp.}} & \multicolumn{1}{c}{$\bolds
{\mathrm{E}(Y)}$} & \multicolumn{1}{c}{$\bolds{\psi}$} &
\multicolumn{1}{c}{\textbf{Bias}} & \multicolumn{1}{c}{\textbf
{ESE}} &
\multicolumn{1}{c}{\textbf{SSE}} & \multicolumn{1}{c}{\textbf
{Cov.}} &
\multicolumn{1}{c}{\textbf{Bias}} & \multicolumn{1}{c}{\textbf
{ESE}} &
\multicolumn{1}{c}{\textbf{SSE}} & \multicolumn{1}{c}{\textbf
{Cov.}} &
\multicolumn{1}{c}{\textbf{Bias}} & \multicolumn{1}{c}{\textbf
{ESE}} &
\multicolumn{1}{c}{\textbf{SSE}} & \multicolumn{1}{c@{}}{\textbf{Cov.}}
\\
\hline
a&0.1&0& 1.62 & 20.1 & 19.6 & 95.6 &-0.10 & 15.9 & 15.6 & 93.7 & 1.23 &
16.4 & 16.1 & 95.5\\
&0.05&1& 5.31 & 33.0 & 32.2 & 96.2&-0.31 & 9.82 & 9.79 & 93.6 & 1.57 &
20.6 & 20.2 & 95.6\\
&0.1&1& 2.71 & 23.6 & 23.0 & 95.6 &-1.01 & 14.2 & 14.3 & 92.6 & 4.00 &
30.1 & 29.5 & 95.6\\
&0.25&1& 1.24 & 15.8 & 15.7 & 95.3 &0.04 & 6.50 & 6.51 & 94.7 & 0.24 &
12.7 & 12.7 & 95.1\\
&0.5&1& 1.46 & 12.6 & 13.0 & 95.6 & 0.32 & 5.44 & 5.56 & 95.9 & 0.29 &
9.93 & 10.3 & 95.9\\[3pt]
b&0.1&0& 2.86 & 28.3 & 28.3 & 95.9 &-0.49 & 15.5 & 15.9 & 94.1 & 1.46 &
15.9 & 16.1 & 95.8\\
&0.05&1& 6.63 & 38.7 & 37.3 & 95.1 &-0.04 & 11.0 & 10.8 & 94.0 & 3.74 &
28.4 & 28.7 & 96.4\\
&0.1&1& 4.37 & 29.0 & 28.9 & 95.2 &0.23 & 8.29 & 8.35 & 94.2 & 2.41 &
20.1 & 21.0 & 96.6\\
&0.25&1& 2.76 & 22.1 & 22.2 & 95.8 &0.18 & 6.42 & 6.49 & 95.1 & 1.24 &
14.2 & 15.1 & 96.5\\
&0.5&1& 1.65 & 19.3 & 19.4 & 95.4 &0.07 & 5.80 & 5.87 & 95.8 & 0.58 &
12.4 & 13.3 & 97.1\\[3pt]
c&0.1&0&5.31 & 39.8 & 39.5 & 95.2&0.85 & 15.7 & 15.7 & 94.8 & 2.47 &
16.5 & 16.5 & 95.3\\
&0.05&1&12.2 & 58.0 & 56.2 & 93.1&1.30 & 17.4 & 17.0 & 91.0 & 7.10 &
36.4 & 36.4 & 93.2\\
&0.1&0&8.15 & 41.1 & 39.7 & 95.9&1.19 & 13.1 & 12.8 & 92.9 & 4.72 &
26.7 & 26.7 & 95.5\\
&0.25&1&3.50 & 26.9 & 25.2 & 95.1&0.35 & 9.24 & 8.69 & 93.9 & 1.62 &
17.2 & 16.8 & 95.6\\
&0.5&1&1.8 & 19.7 & 19.0 & 95.3&0.04 & 6.80 & 6.63 & 95.2 & 0.46 & 12.1
& 12.4 & 96.1\\[3pt]
d&0.1&0&-1.03 & 28.6 & 28.5 & 94.0&0.31 & 21.3 & 20.6 & 92.9 & 2.31 &
21.6 & 21.4 & 97.0\\
&0.05&1&-1.09 & 40.0 & 34.1 & 87.7&-1.52 & 22.0 & 20.0 & 84.3 & 11.3 &
52.4 & 48.9 & 90.0\\
&0.1&0&1.16 & 33.5 & 31.6 & 88.2&0.17 & 14.9 & 14.8 & 90.9 & 6.78 &
36.2 & 35.0 & 93.5\\
&0.25&1&1.59 & 26.8 & 27.0 & 94.0&1.00 & 9.56 & 9.79 & 93.0 & 3.45 &
21.3 & 21.4 & 95.7\\
&0.5&1& -0.07 & 27.3 & 28.5 & 95.2&1.42 & 7.36 & 7.4 & 92.9 & 2.39 &
13.8 & 14.1 & 95.6\\[3pt]
e&0.1&0&3.25 & 26.8 & 26.4 & 97.2&1.66 & 15.2 & 15.2 & 93.4 & -0.73 &
15.1 & 15.1 & 93.9\\
&0.05&1&13.4 & 56.3 & 52.4 & 91.0&6.54 & 41.8 & 36.7 & 82.5 & -1.93 &
13.3 & 11.5 & 85.2\\
&0.1&0&8.38 & 39.8 & 38.0 & 93.9&6.50 & 33.5 & 31.8 & 87.2 & -0.38 &
11.1 & 10.7 & 83.6\\
&0.25&1&4.83 & 24.4 & 24.3 & 95.7&2.88 & 22.1 & 22.4 & 93.3 & 0.46 &
9.51 & 9.64 & 91.9\\
&0.5&1&4.18 & 17.1 & 17.4 & 96.0&-3.39 & 19.3 & 19.9 & 94.4 & 0.08 &
11.3 & 11.9 & 95.1\\
\hline
\end{tabular*}
\vspace*{-3pt}
\end{table*}

%s5 ###
\section{Applications}\label{dataanalysis}

%s5.1 ###
\subsection{Analysis of a Health Register}\label{dataanalysisbrookhart}

Brookhart et al. (\citeyear{Broetal06}) and Brookhart and Schnee\-weiss
(\citeyear{BroSch07}) assess
short-term effects of Cox-2 treatment (as compared to nonsteroidal\vadjust{\goodbreak}
anti-inflammato\-ry treatment) on the risk of gastrointestinal (GI)
bleeding within 60 days. As Table \ref{tab4} shows, of the 37,842 new
nonselective NSAID users drawn from a~large population based cohort of
medicare beneficiaries who were eligible for a state-run pharmaceutical
benefit plan, 26,407 patients were placed on \mbox{Cox-2} treatment.
Let the received treatment $X$ equal $1$ for subjects placed on Cox-2
and $0$ for those on nonselective NSAIDs. Let the outcome $Y$ indicate~1
for upper gastrointestinal (GI) bleeding within 60 days of initiating
an NSAID and 0 otherwise. As in Brookhart and Schneeweiss (\citeyear
{BroSch07}), we
use the physician's prescribing preference for Cox-2 (versus
nonselective NSAIDs) $Z$ as an instrumental variable for the effect of
Cox-2 treatment on gastrointestinal bleeding.
The Wald and adjusted IV estimator of the conditional causal odds ratio
were found to be identical: 0.26 (95\% confidence interval 0.084--0.79,
P~0.018). In contrast, the logistic structural mean model
estimator [both using the approach of Vansteelandt and Goetghebeur\vadjust{\goodbreak}
(\citeyear{VanGoe03}) and using
the approach of Robins and Rotnitzky (\citeyear{RobRot04})] was found
to be 0.081 (95\% confidence interval 0.0095--0.82, P~0.018), which
might be more reliable, considering the nonnormality of the exposure
distribution. The marginal causal odds ratio was estimated to be almost
identical: 0.083 (95\% confidence interval 0.0096--0.82). We thus
estimate roughly that the use of nonselective NSAIDs instead of Cox-2
increases the odds (or risk) of gastrointestinal bleeding by at least
18\% ($=1-0.82$).

Besides the IV assumptions, all results rely on the assumption that the
effect of Cox-2 versus nonselective NSAIDS is the same in Cox-2 users
whose physician prefers Cox-2 treatment as in Cox-2 users whose
physician prefers nonselective NSAIDS (and likewise for the effect of
nonselective NSAIDS). They are in stark contrast with the estimate
obtained from an unadjusted logistic regression analysis: 1.12 (95\%
confidence interval 0.85--1.5).

%s5.2 ###
\subsection{Analysis of Randomized Cholesterol Reduction Trial with
Noncompliance}

We reanalyze the cholesterol reduction trial reported in Ten Have et
al. (\citeyear{HavJofCar03}). Let $Y$ be an indicator of treatment
success (defined as a
beneficial change in cholesterol), $X$ be an indicator of using
educational dietary home-based audio tapes (which equals 0 on the
control arm) and $Z$ be the experimental assignment to the use of
educational dietary home-based audio tapes.
The Wald estimator of the conditional causal odds ratio was found to be
1.37 (95\% confidence interval 0.68--2.74, P~0.38), and analogous to
the logistic structural mean model estimator, 1.31 (95\% confidence
interval 0.72--2.40, P~0.37). This expresses that in patients who
used the audio tapes on the intervention arm, the odds of a beneficial
reduction in cholesterol would have been 1.31 times lower had they not
received the intervention. The adjusted IV estimator was uninformative:
0.020 (95\% confidence interval 0--10$^{171}$, P~0.99).
%
%t4 ###
%
\begin{table}
\tabcolsep=0pt
\caption{Observed data with $X_i$ indicating received treatment [Cox-2
(1) versus nonselective NSAIDs (0)], $Z_i$ indicating the physician's
prescribing preference [Cox-2 (1) versus nonselective NSAIDs (0)], and
$Y_i$ indicating gastrointestinal (GI) bleeding (1) within 60 days of
initiating an NSAID for subject $i$}\label{tab4}
\begin{tabular*}{\columnwidth}{@{\extracolsep{\fill}}lcccc@{}}
\hline
& \multicolumn{2}{c}{$\bolds{Z_i=0}$} & \multicolumn{2}{c@{}}{$\bolds
{Z_i=1}$}\\
\ccline{2-3,4-5}\\[-8pt]
&$\bolds{Y_i=0}$ & $\bolds{Y_i=1}$& $\bolds{Y_i=0}$ & $\bolds{Y_i=1}$\\
\hline
$X_i=0$ & 5640 & 39 & \phantom{0}5722 & \phantom{0}34 \\
$X_i=1$ & 6740 & 60 & 19493 & 114 \\
\hline
\end{tabular*}
\end{table}
The marginal causal odds ratio (\ref{or1}) was estimated to be 1.28
(95\% confidence interval 0.74--2.19, P~0.38). It expresses that, had
all patients complied perfectly with their assigned treatment, the
intention-to-treat analysis would have resulted in an odds ratio of
1.28. Since the exposure is dichotomous, the marginal causal odds ratio
(\ref{or2}) is not of interest. Since subjects on the control arm have
no access to the audio tapes, model (\ref{VGmodelsimple}) is only
relevant for those who were assigned to the intervention arm (i.e.,
$Z=1$); hence, this analysis does not rely on untestable assumptions
regarding the absence of exposure effect modification by the
instrumental variable.

%s5.3 ###
\subsection{Analysis of Randomized Blood Pressure Trial With Noncompliance}

We reanalyze the blood pressure study reported in Vansteelandt and
Goetghebeur (\citeyear{VanGoe03}). Let $Y$ be an indicator of
successful blood
pressure reduction,~$X$ measure the percentage of assigned active dose
which was actually taken (which equals 0 on the control arm) and $Z$ be
the experimental assignment to active treatment or placebo.
The Wald and adjusted IV estimator of the conditional causal odds ratio
were found to be identical, 4.29 (95\% confidence interval 1.6--11.3,
P~0.0032), and analogous to the logistic structural mean model
estimator, 4.44 (95\% confidence interval 1.6--12.6, P~0.0049). This
expres\-ses that in patients on the intervention arm with unit exposure
per day, the odds of a beneficial reduction in diastolic blood pressure
would have been 4.44 times lower had they not received the experimental
treatment. The marginal causal odds ratio~(\ref{or1}) was estimated to
be 4.12 (95\% confidence interval 1.6--10.3, P~0.0025). It expresses
that, had all patients complied perfectly with their assigned
treatment, the intention-to-treat analysis would have resulted in an
odds ratio of 4.12.

\appendix

%s6 ###
\section*{Appendix}\label{app}

%s6.1 ###
\setcounter{subsection}{0}
\subsection{Closed-Form Estimator}

When $X$ and $Z$ are both dichotomous, taking values 0 and 1, the
logistic structural mean model estimator is obtainable in closed form as
\renewcommand{\theequation}{\arabic{equation}}
\setcounter{equation}{29}
\begin{equation}%\label{closedform}
\hat{\psi}=\log\biggl[\frac{-Q_1\pm\sqrt{Q_1^2-4 Q_2 (Q_2-\hat
{X}_{11}+\hat{X}_{10})Q_3}}{2 Q_2}\biggr],
\end{equation}
where $\hat{X}_{xz}$ is the percentage of subjects with $X=x$ among
those with $Z=z$, and
\begin{eqnarray*}
Q_1&=&(Q_2+\hat{X}_{10})\exp(\hat{\beta}_0+\hat{\beta
}_1)\\
&&{}+(Q_2-\hat
{X}_{11})\exp(\hat{\beta}_0+\hat{\beta}_1+\hat{\beta}_2+\hat
{\beta}_3),\\
Q_2&=&\expit(\hat{\beta}_0)\hat{X}_{00}-\expit(\hat{\beta
}_0+\hat{\beta
}_2)\hat{X}_{01},\\
Q_3&=&\exp(\hat{\beta}_0+\hat{\beta}_1+\hat{\beta}_2+\hat{\beta
}_3)\times\exp
(\hat{\beta}_0+\hat{\beta}_1).
\end{eqnarray*}

%s6.2 ###
\subsection{Standard Errors for Conditional Causal Log Odds Ratio Estimators}

Suppose that $X$ satisfies the conditional mean mo\-del
\[
\mathrm{E}(X|Z,C)=g(Z,C;\theta^*),
\]
where $g(Z,C;\theta)$ is a known function, smooth in $\theta$, and
$\theta^*$ is an unknown finite-dimensional parameter; for example,
$g(Z,C;\theta)=\theta_0+\theta_1Z+\theta_2 C$.
With $R(\theta^*)\equiv X-g(Z,C;\theta^*)$, assume further that
\begin{eqnarray*}
&&\logit\mathrm{E}(Y|Z,C,R(\theta^*))\\
&&\quad=m_0(C,R(\theta^*);\omega
^*)+m(C;\psi
^*)g(Z,C;\theta^*),
\end{eqnarray*}
where $m_0(C,R(\theta^*);\omega)$ is a known function, smooth in
$\omega
$, and $\omega^*$ is an unknown finite-dimensional parameter; for
example, $m_0(C,R(\theta^*);\omega)=\omega_0+\omega_1 C+\omega_2
R(\theta^*)$.
Then the adjusted IV estimator is equivalently obtained by solving the
multivariate score equation\vadjust{\goodbreak} $\sum^{n}_{i=1}S_{i}(\xi) = {0}$ for $\xi
\equiv(\theta',\omega',\psi')'$ and taking the solution for $\psi$,
where $S_{i}(\theta,\omega,\psi)$ equals
%
%e18 ###
%
\renewcommand{\theequation}{\arabic{equation}}
\setcounter{equation}{30}
\begin{equation}
\qquad\pmatrix{
{\displaystyle\frac{\partial g}{\partial\theta}(Z_i,C_i;\theta
)\operatorname{Var}^{-1}(X_i|Z_i,C_i)R_i(\theta)}\vspace*{2pt}\cr
\hspace*{-46pt}\pmatrix{
\displaystyle\frac{\partial m_0}{\partial\omega}(C_i,R_i(\theta);\omega
)\vspace*{2pt}\cr
\displaystyle\frac{\partial m}{\partial\psi}(C_i;\psi)g(Z_i,C_i;\theta
)}\vspace*{4pt}\cr
\hspace*{-14pt}{}\cdot[Y_i-\expit\{m_0(C_i,R_i(\theta);\omega)\vspace
*{2pt}\cr
\hspace*{16pt}{}+m(C_i;\psi
)\cr
\hspace*{70pt}\cdot{} g(Z_i,C_i;\theta)\}]}.
\end{equation}
The asymptotic variance of the adjusted IV estimator can now be
obtained from the ``sandwich'' expression
\[
\frac{1}{n}\mathrm{E}^{-1}\biggl(\frac{\partial S_{i}(\xi)}{\partial\xi}
\biggr)\operatorname{Var}\{S_{i}(\xi)\}\mathrm{E}^{-1}\biggl(\frac
{\partial
S_{i}(\xi)}{\partial\xi}\biggr)^{T}.
\]
The asymptotic variance of the standard IV estima\-tor is similarly
obtained upon redefining $m_0(C,R(\theta^*);\allowbreak \omega)$ to be a function
of only $C$ and $\omega$. The asymptotic variance of the logistic
SMM-estimator is obtained as in Vansteelandt and Goetghebeur (\citeyear
{VanGoe03}).

%s6.3 ###
\subsection{Theoretical Comparison of the Adjusted IV Estimator and the
Logistic Structural Mean Model Estimator}

To simplify the exposition, suppose that there are no covariates.
Assume that $X$ is normally distributed, conditional on $Z$. Let the
adjusted IV estimator be based on the model
\[
\operatorname{logit} \mathrm{P}(Y=1|R,Z)=\omega_0+\omega_1 R +\omega_2
\mathrm{E}(X|Z),
\]
and assume, for the purpose of comparability, that this is also the
association model underlying the logistic structural mean model
estimator [e.g., when $\mathrm{E}(X|Z)$ is linear in $Z$, then this is
equivalent with a~standard logistic regression model with main effects
in $X$ and $Z$]. Under model (\ref{VGmodelsimple}), it then follows that
\begin{eqnarray*}
&&\operatorname{logit} \mathrm{P}\bigl(Y(0)=1|X,Z\bigr)\\
&&\quad=\omega_0+(\omega_1-\psi)R+(\omega
_2-\psi)\mathrm{E}(X|Z).
\end{eqnarray*}
We will now demonstrate that the adjusted IV estimator $\hat{\omega}_2$
is a consistent estimator of the causal parameter $\psi^*$ indexing the
logistic structural mean model. We will do so by demonstrating that the
estimating equations for the logistic structural mean model estimator
$\hat{\psi}$ have mean zero at $\psi=\omega_2$.

Note that, at $\omega_2=\psi$, $\operatorname{logit} \mathrm
{P}(Y(0)=1|X,Z)=\omega
_0+(\omega_1-\psi)R$.\vadjust{\goodbreak}
A Taylor series expansion of the estimating function for $\psi$, that is,
\[
[d(Z)\,{-}\,\mathrm{E}\{d(Z)\}]\operatorname{expit}[\omega_0\,{+}\,(\omega_1\,{-}\,\psi)\{X\,{-}\,\mathrm{E}(X|Z)\}],
\]
around $X=\mathrm{E}(X|Z)$ then gives
\begin{eqnarray*}
&&\sum_{k=0}^{\infty}[d(Z)-\mathrm{E}\{d(Z)\}]\{X-\mathrm{E}(X|Z)\}^k\\
&&\hspace*{14pt}{}\cdot\operatorname{expit}^{(k)}(\omega_0)\frac{(\omega
_1-\psi)^k}{k!},
\end{eqnarray*}
where $\operatorname{expit}^{(k)}(\omega_0)$ refers to the $k$th
order derivative of $\operatorname{expit}(\omega_0)$ w.r.t. $\omega_0$.
When $X$ is normally distributed, conditional on $Z$, with constant
variance, then this is a mean zero equation because then $\mathrm{E}[\{
X-\mathrm{E}(X|Z)\}^k|Z]=\mathrm{E}[\{X-\mathrm{E}(X|Z)\}^k]$ for
all~$k$. It
thus follows that $\hat{\omega}_2$ is a consistent estimator of the
causal parameter $\psi^*$. This result continues to hold for other
distributions than the normal, which satisfy that for each $k$, either
$\mathrm{E}[\{X-\mathrm{E}(X|Z)\}^k|Z]=\mathrm{E}[\{X-\mathrm{E}(X|Z)\}
^k]$ or $\operatorname{expit}^{(k)}(\omega_0)=0$. For instance, when
$X$ is
normally distributed, conditional on $Z$, with variance depending on
$Z$ and when, in addition, $\operatorname{expit}(\omega_0)=1/2$,
then $\hat{\omega}_2$ stays a consistent estimator of the causal\vspace*{1.2pt}
parameter $\psi^*$ because $\mathrm{E}[\{X-\mathrm{E}(X|Z)\}
^k|Z]=\mathrm{E}[\{
X-\mathrm{E}(X|Z)\}^k]$ for all $k\ne2$ and $\operatorname{expit}^{(2)}
(\omega_0)=\operatorname{expit}(\omega_0)\{1-\operatorname{expit}(\omega
_0)\}\{1-\break 2\operatorname{expit}(\omega_0)\}=0$.

%s6.4 ###
\subsection{Local Robustness}

Suppose first that $C_i$ is empty, $d(Z_i,C_i)=Z_i$ and $\mathrm{E}\{
d(Z_i,C_i)|C_i\}=\sum_{j=1}^n Z_j/n$. When $\psi^*=0$, then
equation (\ref{VGestimatingequation}) becomes
$\sum_{i=1}^n (Z_i-\frac{\sum_{j=1}^n Z_j}{n})\cdot\break\expit\{
m(X_i,Z_i; \hat{\beta})\}$.
Suppose now that the association model includes an intercept and main
effect in~$Z_i$, and that $\hat{\beta}$ is the standard maximum
likelihood estimator of $\beta^*$. We then show that equation (\ref
{VGestimatingequation}) equals
$\sum_{i=1}^n (Z_i-\frac{\sum_{j=1}^n Z_j}{n})Y_i$, which
has mean zero at $\psi^*=0$, even under model misspecification. That
this equality is true follows because $\hat{\beta}$ satisfies the
following score equations:
\[
0=\sum_{i=1}^n\pmatrix{ 1\cr Z_i}
[Y_i-\expit\{m(X_i,Z_i; \hat{\beta})\}]
\]
from which $\sum_{i=1}^n Z_iY_i=\sum_{i=1}^n Z_i\expit\{m(X_i,Z_i;
\hat
{\beta})\}$ and
\[
\sum_{i=1}^n \frac{\sum_{j=1}^n Z_j}{n}Y_i=\sum_{i=1}^n \frac{\sum
_{j=1}^n Z_j}{n}\expit\{m(X_i,Z_i; \hat{\beta})\}.
\]
Extending this argument, it is seen that local robustness is attained
whenever the association model includes an additive term in
$d(Z_i,C_i)-\mathrm{E}\{d(Z_i,C_i)|C_i\}$.

%s6.5 ###
\subsection{Uncongenial Models}

It follows from the parameterization of Robins and Rotnitzky (\citeyear
{RobRot04})
that, for each law $f(X|Z,C)$, the logistic structural mean model (\ref
{VGmodel}) is congenial with association models of the form
\begin{eqnarray*}
&&\mathrm{P}(Y=1|X,Z,C)\\
&&\quad=\expit\{m(C;\psi^*) X+ q(X,Z,C)+v(Z,C)\}
\end{eqnarray*}
for each function $q(X,Z,C)$ of $(X,Z,C)$ satisfying $q(0,Z,C)=0$ for
all $Z,C$, each function $t(C)$ of $C$, and $v(Z,C)$ solving
\begin{eqnarray*}
t(C)&=& \int\expit\{q(X=x,Z,C)+v(Z,C)\}\\
&&\hspace*{9pt}{}\cdot f(X=x|Z,C)\,dx.
\end{eqnarray*}
It thus follows that, for each law $f(X|Z,C)$, the logistic structural
mean model (\ref{VGmodel}) is also congenial with association models
of the form
%
%e19 ###
%
\begin{eqnarray}\label{assocnew}
&&\mathrm{P}(Y=1|X,Z,C)\nonumber\\
&&\quad=\expit\{m(C;\psi^*) X+ q(X,Z,C)\\
&&\hspace*{50pt}\qquad{}+t^*(C)+v^*(Z,C)
\}\nonumber
\end{eqnarray}
for each such function, each function $t^*(C)$ of $C$, and $v^*(Z,C)$
satisfying $v^*(0,C)=0$ for all $C$ and
%
%e21 ###
%e20 ###
%
\begin{eqnarray}\label{intnew}
&&\int\expit\{q(X=x,0,C)+t^*(C)\}\nonumber\\
&&\quad\cdot f(X=x|Z=0,C)\,dx
\nonumber\\
&&\quad= \int
\expit\{
q(X=x,Z,C)\\
&&\hspace*{36pt}\qquad{}+t^*(C)+v^*(Z,C)\}\nonumber\\
&&\hspace*{34pt}\cdot f(X=x|Z,C)\,dx\nonumber
\end{eqnarray}
for each $Z$. Indeed, this follows upon defining $t^*(C)$ as the
solution to
\begin{eqnarray*}
&&t(C)=\int\expit\{q(X=x,0,C)+t^*(C)\}\\
&&\hspace*{45pt}\cdot f(X=x|Z=0,C)\,dx.
\end{eqnarray*}
It follows that a given association model is congenial with the
logistic structural mean model (\ref{VGmodel}) when no restrictions
are imposed on the function $v^*(Z,C)$, which encodes the main effect
of $Z$, along with interactions with $C$. The above derivation also
suggests an easier strategy for fitting the model of Robins and
Rotnitzky (\citeyear{RobRot04}), whereby the association model is of
the form (\ref
{assocnew}) and integral equations of the form~(\ref{intnew}) are
solved.

Consider now the extended logistic SMM (\ref{GVGmodel}).
Suppose that model (\ref{GVGmodel}) is congenial with the association
model (\ref{association}) for $x=0$ in the\vadjust{\goodbreak} sense that for
the given $\beta^*$, there exists a value $\psi_0^*$ such that
%each $\psi
%_0^*$, there exists a value $\beta^*$ such that (\ref{GVGmodel}) and
%(\ref{association}) hold for $x=0$ and that
%
\[
\int\expit\{m(X,Z,C;\beta^*)-m(C;\psi^*_0) X\}f(X|Z,C)\,dX
\]
does not depend on $Z$. Then it does not necessarily follow that there
exists a value $\psi_x^*$ for given $x$ such that
\begin{eqnarray*}
&&\int\expit\{m(X,Z,C;\beta^*)\\
&&\hspace*{34pt}{}-m(C;\psi^*_x) (X-x)\}f(X|Z,C)\,dX
\end{eqnarray*}
does not depend on $Z$. Model (\ref{GVGmodel}) being congenial with
the association model (\ref{association}) for $x=0$ hence does not
imply congeniality for all $x$.

%s6.6 ###
\subsection{Probit-Normal SMM Estimator}

We explain how to derive $\mathrm{E}(Y(0)|Z,C)$ under models (\ref{probitsmm})
and (\ref{assoc-probit}). Note that
\begin{eqnarray*}
&&\mathrm{E}\{Y(0)|Z,X,C\}\\
&&\quad= \mathrm{P}(U \leq\theta^*_0+\theta^*_1 X+\theta
^*_2 Z+\theta^*_3C-\phi^* X),
\end{eqnarray*}
where $U$ is a standard normally distributed variate, independent of
$(Z,X)$. Averaging over the exposure, conditional on $Z$ and $C$, then yields
\begin{eqnarray*}
&&\mathrm{E}\{Y(0)|Z,C\}\\
&&\quad=\int_{-\infty}^\infty\mathrm{P}\bigl(U+(\phi^* -\theta
^*_1) X\\
&&\qquad\hspace*{35pt}\leq\theta^*_0+\theta^*_2 Z+\theta^*_3C\bigr)
\,dF(X|Z,C),
\end{eqnarray*}
where $F(X|Z,C)$ refers to the conditional distribution of $X$, given
$Z$ and $C$.
Define $U^*=U+(\phi^* -\theta^*_1) X$. Then, for normally distributed
$X$ with mean $\alpha^*_0+\alpha^*_1 Z+\alpha_2^*C$ and constant
variance $\sigma^{2*}$, conditional on $Z$ and $C$, $U^*$ has a normal
distribution with mean $(\phi^*-\theta_1^*)(\alpha_0+\alpha_1
Z+\alpha
_2^*C)$ and variance $1+(\phi^*-\theta_1^*)^2 \sigma^2$. Then
\[
\mathrm{E}\{Y(0)|Z,C\}=\int_{-\infty}^\infty\int_{-\infty
}^{\theta
_0^*+\theta_2^* Z+\theta_3^*C} \,dF(U^*, X|Z,C),
\]
which is as given in (\ref{cumulative}). The conditional mean $\mathrm{E}
(Y|Z,C)$ can be derived using similar arguments.

%s6.7 ###
\subsection{Standard Errors for Marginal Causal Log Odds Ratio Estimators}

Consider the marginal log odds ratio defined by
%
%e22 ###
%
\begin{equation}\label{eqmlor}
\eta=\log\frac{\mu_1(1-\mu_0)}{\mu_0(1-\mu_1)},
\end{equation}
where $\mu_x=\mathrm{E}[\operatorname{expit}\{m(X,Z,C;\beta^*)+m(C;\psi
^*_{x})(x-X)\}]$ for $x=0,1$, and let the corresponding
estimators be $\hat{\eta}$ and $\hat{\mu}_x,x=0,1$,\vadjust{\goodbreak} respectively. Then
a~Taylor series expansion shows that
\begin{eqnarray*}
0&=&\frac{1}{\sqrt{n}}\sum_{i=1}^n \operatorname{expit}\{
m(X_i,Z_i,C_i;\hat{\beta})\\
&&\hspace*{58pt}{}+m(C_i;\hat{\psi}_{x})(x-X_i)\}-\hat{\mu
}_x\\
&=& \frac{1}{\sqrt{n}}\sum_{i=1}^n\operatorname{expit}\{
m(X_i,Z_i,C_i;\beta)\\
&&\hspace*{58pt}{}+m(C_i;\psi_{x})(x-X_i)\}-{\mu}_x\\
&&{}+\frac{1}{\sqrt{n}}\sum_{i=1}^n\mathrm{E}\biggl[\frac{\partial}{\partial
\theta
_x}\operatorname{expit}\{m(X_i,Z_i,C_i;\beta)\\
&&\hspace*{82pt}\qquad{}+m(C_i;\psi
_{x})(x-X_i)\}\biggr]\\
&&\hspace*{46pt}{}\cdot
\mathrm{E}^{-1}\biggl(\frac{\partial U_{ix}(\theta_x)}{\partial\theta_x}
\biggr)U_{ix}(\theta_x)\\
&&\hspace*{46pt}{}-\sqrt{n}(\hat{\mu}_x-\mu_x),
\end{eqnarray*}
where $\theta_x\equiv(\beta^T,\psi_x^T)^T$ and $U_{ix}(\theta_x)$ is
the vector of estimating functions for $\theta_x$,
from which the influence function for $\hat{\mu}_x$ is
\begin{eqnarray*}
&&\operatorname{expit}\{m(X_i,Z_i,C_i;\beta)+m(C_i;\psi_{x})(x-X_i)
\}-{\mu}_x\\
&&\quad{}+\frac{1}{\sqrt{n}}\sum_{i=1}^n\mathrm{E}\biggl[\frac{\partial}{\partial
\theta
_x}\operatorname{expit}\{m(X_i,Z_i,C_i;\beta)\\
&&\hspace*{116pt}{}+m(C_i;\psi
_{x})(x-X_i)\}\biggr]\\
&&\hspace*{56pt}{}\cdot
\mathrm{E}^{-1}\biggl(\frac{\partial U_{ix}(\theta_x)}{\partial\theta_x}
\biggr)U_{ix}(\theta_x).
\end{eqnarray*}
From the Delta method, it then follows that the influence function for
$\hat{\eta}$ is
\begin{eqnarray*}
&& \frac{1}{\mu_1(1-\mu_1)}\\
&&\quad\cdot\Biggl[\operatorname{expit}\{
m(X_i,Z_i,C_i;\beta)\\
&&\hspace*{44pt}{}+m(C_i;\psi_{1})(1-X_i)\}-{\mu}_1\\
&&\hspace*{18pt}{}+\frac{1}{\sqrt{n}}\sum_{i=1}^n\mathrm{E}\biggl[\frac{\partial
}{\partial
\theta_{1}}\operatorname{expit}\{m(X_i,Z_i,C_i;\beta)\\
&&\hspace*{122pt}{}+m(C_i;\psi
_{1})(1-X_i)\}\biggr]\\
&&\hspace*{112pt}{}\cdot
\mathrm{E}^{-1}\biggl(\frac{\partial U_{i1}(\theta_{1})}{\partial\theta
_{1}}\biggr)U_{i1}(\theta_{1})\Biggr]\\
&&\quad{}-\frac{1}{\mu_0(1-\mu_0)}\\
&&\qquad{}\cdot\Biggl[\operatorname{expit}\{
m(X_i,Z_i,C_i;\beta)\\
&&\hspace*{60pt}{} +m(C_i;\psi_{0})(0-X_i)\}-{\mu}_0\\
&&\hspace*{34pt}{}+\frac{1}{\sqrt{n}}\sum_{i=1}^n\mathrm{E}\biggl[\frac{\partial
}{\partial
\theta_0}\operatorname{expit}\{m(X_i,Z_i,C_i;\beta)\\
&&\hspace*{108pt}\qquad{}+m(C_i;\psi
_{0})(0-X_i)\}\biggr]\\
&&\hspace*{114pt}{}\cdot
\mathrm{E}^{-1}\biggl(\frac{\partial U_{i0}(\theta_0)}{\partial\theta_0}
\biggr)U_{i0}(\theta_0)\Biggr].
\end{eqnarray*}
The asymptotic variance of $\hat{\eta}$ thus equals 1 over $n$ times
the variance of this influence function (where averages and variances
can be replaced with sample analogs, and population values with
consistent estimators).

Consider the marginal log odds ratio defined by~(\ref{eqmlor}) with
the redefinitions
\[
\mu_1=\mathrm{E}[\operatorname{expit}\{m(X,Z,C;\beta^*)+m(C;\psi
^*_{X+1})\}]
\]
and $\mu_0=\mathrm{E}(Y)$. Then using similar arguments as before, we obtain
that the influence function for $\hat{\eta}$ is
\begin{eqnarray*}
&& \frac{1}{\mu_1(1-\mu_1)}\\
&&\quad{}\cdot\Biggl[\operatorname{expit}\{
m(X_i,Z_i,C_i;\beta)\\
&&\hspace*{48pt}{}+m(C_i;\psi_{X_i+1})\}-{\mu}_1\\
&&\hspace*{22pt}{}+\frac{1}{\sqrt{n}}\sum_{i=1}^n\mathrm{E}\biggl[\frac{\partial
}{\partial
\theta_{X_i+1}}\operatorname{expit}\{m(X_i,Z_i,C_i;\beta)\\
&&\hspace*{120pt}\qquad{}+m(C_i;\psi
_{X_i+1})\}\biggr]
\\
&&\hspace*{48pt}{}\cdot\mathrm{E}^{-1}\biggl(\frac{\partial
U_{i,X_i+1}(\theta
_{X_i+1})}{\partial\theta_{X_i+1}}\biggr)U_{i,X_i+1}(\theta
_{X_i+1})\Biggr] \\
&&\quad{}-\frac{1}{\mu_0(1-\mu_0)}[Y_i-{\mu}_0].
\end{eqnarray*}

\section*{Acknowledgments}

We are grateful to Alan Brookhart for providing the data, to Vanessa
Didelez and Tom Palmer for helpful discussions, and to three referees whose
comments substantially improved an earlier version of this article. We
acknowledge support from\vadjust{\goodbreak} Ghent University (Multidisciplinary Research
Partnership ``Bioinformatics: from nucleotides to networks'') and IAP
research network Grant nr. P06/03 from the Belgian government (Belgian
Science Policy). The second author was supported by MRC Grant nr.
U1052.00.014. The third author would like to thank Iran's Minister
of Science for financially supporting his PhD study at Ghent
University. The fourth author acknowledges partial support from NIH
Grant AI24643.

% imsref loaded by akundreckaite, 2011-06-16 12:13:58
%
% imsref loaded by akundreckaite, 2011-06-16 12:20:10
%

\end{document}